\documentclass{aa}
\usepackage[varg]{txfonts}
\usepackage{chemformula}
\usepackage{graphicx}
\usepackage[colorlinks=true, linkcolor=blue, citecolor=blue, filecolor=blue, urlcolor=blue]{hyperref}
                        
\begin{document}

\title{The GAPS Programme at TNG\thanks{Based on observations made with the Italian Telescopio Nazionale Galileo (TNG) operated by the Fundaci\'{o}n Galileo Galilei (FGG) of the Istituto Nazionale di Astrofisica (INAF) at the Observatorio del Roque de los Muchachos (La Palma, Canary Islands, Spain).}}
\subtitle{LXVII. Detection of water and a preliminary characterisation of the atmospheres of the two hot Jupiters: KELT-8\,b and KELT-23\,Ab}

\author{
M. Basilicata\inst{1,2}
\and 
P. Giacobbe\inst{2}
\and
M. Brogi\inst{3, 2}
\and
F. Amadori\inst{2}
\and
E. Pacetti\inst{4}
\and
M. Baratella\inst{5}
\and
A. S. Bonomo\inst{2}
\and
K. Biazzo\inst{6}
\and
D. Turrini\inst{2}
\and
L. Mancini\inst{1,2,7}
\and
A. Sozzetti\inst{2}
\and
G. Andreuzzi\inst{8,6}
\and
W. Boschin\inst{8,9,10}
\and
L. Cabona\inst{11}
\and
S. Colombo\inst{12}
\and
M. C. D'Arpa\inst{12,13}
\and
G. Guilluy\inst{2}
\and
A. F. Lanza\inst{14}
\and
L. Malavolta\inst{15,16}
\and
F. Manni\inst{1,2}
\and
L. Naponiello\inst{2}
\and
M. Pinamonti\inst{2}
\and
L. Pino\inst{17}
\and
D. Sicilia\inst{14}
\and
T. Zingales\inst{15,16}
}
\institute{
Department of Physics, University of Rome ``Tor Vergata'', Via della Ricerca Scientifica 1, I-00133 Roma, Italy \\
\email{mario.basilicata@inaf.it}
\and 
INAF -- Osservatorio Astrofisico di Torino, Via Osservatorio 20, I-10025 Pino Torinese, Italy
\and
Department of Physics, University of Turin, Via Pietro Giuria 1, I-10125 Torino, Italy
\and
Istituto di Astrofisica e Planetologia Spaziali INAF-IAPS, Via Fosso del Cavaliere 100, I-00133 Roma, Italy
\and
ESO -- European Southern Observatory, Alonso de Cordova, 3107 Vitacura, Santiago, Chile
\and
INAF -- Osservatorio Astronomico di Roma, Via Frascati 33, I-00078 Monte Porzio Catone, Italy
\and
Max Planck Institute for Astronomy, K\"{o}nigstuhl 17, I-69117 Heidelberg, Germany
\and
Fundaci\'{o}n Galileo Galilei -- INAF, Rambla Jos\'{e} Ana Fernandez P\'{e}rez 7, E-38712 -- Bre\~{n}a Baja (La Palma), Canary Islands, Spain
\and
Instituto de Astrof\'{\i}sica de Canarias, C/V\'{\i}a L\'actea s/n, E-38205 La Laguna (Tenerife), Canary Islands, Spain
\and
Departamento de Astrof\'{\i}sica, Univ. de La Laguna, Av. del Astrof\'{\i}sico Francisco S\'anchez s/n, E-38205 La Laguna (Tenerife), Canary Islands, Spain
\and
INAF -- Osservatorio Astronomico di Brera, Via E. Bianchi 46, I-23807 Merate, Italy
\and
INAF -- Osservatorio Astronomico di Palermo, Piazza del Parlamento 1, I-90134 Palermo, Italy
\and
Department of Physics and Chemistry, University of Palermo, Piazza Marina 61, I-90133 Palermo, Italy
\and
INAF -- Osservatorio Astrofisico di Catania, Via S. Sofia 78, I-95123 Catania, Italy
\and
Department of Physics and Astronomy Galileo Galilei, University of Padua, Vicolo dell'Osservatorio 3, I-35122 Padova, Italy
\and
INAF -- Osservatorio Astronomico di Padova, Vicolo dell’Osservatorio 5, I-35122 Padova, Italy
\and
INAF -- Osservatorio Astrofisico di Arcetri, Largo Enrico Fermi 5, I-50125 Firenze, Italy
}

\abstract{Hot Jupiters are among the most suitable targets for atmospheric studies. Expanding the number of hot gaseous giant planets with atmospheric characterisations can improve our understanding of the chemical-physical properties of their atmospheres, as well as the formation and evolution of these extreme planets.}
{In this work, we use high-resolution spectroscopy in the near-infrared (NIR) to search for chemical signatures in the atmosphere of the two hot Jupiters KELT-8\,b ($T_{\rm eq}=1\,675^{+61}_{-55}$\,K) and KELT-23\,Ab ($T_{\rm eq}=1\,561\pm20$\,K) and present a first characterisation of their atmospheric properties.}
{We measured the transmission spectrum of each target with the near-infrared (NIR) high-resolution spectrograph GIANO-B at the TNG. We searched for atmospheric signals by cross-correlating the data with synthetic transmission spectra. To characterise the chemical-physical properties of the atmosphere of both planets, we ran two different atmospheric retrievals for each dataset: a retrieval assuming chemical equilibrium and a `free-chemistry' retrieval, in which the abundance of each molecule could vary freely.}
{We detect water vapour (\ch{H2O}) in the atmospheres of both KELT-8\,b and KELT-23\,Ab with a signal-to-noise ratio of S/N = 6.6 and S/N = 4.2, respectively. The two retrievals indicate a water-rich atmosphere for both targets. In the case of KELT-8\,b, we determine a water volume mixing ratio of log$_{10}$(VMR$_{\rm H_2O})=-2.07^{+0.53}_{-0.72}$, a metallicity of [M/H] $=0.77^{+0.61}_{-0.89}$\,dex, and a sub-solar C/O ratio (C/O $\leq0.30$, at $2\,\sigma$). For KELT-23\,Ab, we find log$_{10}$(VMR$_{\rm H_2O})=-2.26^{+0.75}_{-1.24}$, [M/H] $=-0.42^{+1.56}_{-1.35}$\,dex, and C/O ratio $\leq0.78$ (at $2\,\sigma$). The constraints on the metallicity and C/O ratio are based on the assumption of chemical equilibrium. Comparing these atmospheric chemical properties with those of the host stars, we suggest that for both planets, the accretion of gaseous material occurred within the \ch{H2O} snowline in a pebble-rich disk enriched in oxygen due to sublimation of water ice from the inward-drifting pebbles.}
{We investigated the atmosphere of KELT-8\,b and KELT-23\,Ab for the first time, finding water vapour in both of them and placing first constraints on their properties. These two planets are promising targets for future high- and low-resolution observations.}
\maketitle

\section{Introduction}
\label{introduction}
The study of exoplanetary atmospheres plays a crucial role in the exoplanet characterisation process. For example, the study of relative abundances of atmospheric species can help to constrain the formation and evolution paths experienced by an exoplanet (e.g.~\citealt{oberg2011,madhusudhan2014,mordasini2016,madhusudhan2019,banzatti2020,bitsch2022,pacetti2022}). Moreover, by studying the chemical properties of exo-atmospheres, we can place stronger constraints on the internal composition of planets, breaking possible degeneracies from the measurement of the bulk density alone (e.g.~\citealt{madhusudhan2020}). The exo-atmosphere investigation also offers the opportunity to study environments without analogues in the Solar System. This is the case, for example, for hot Jupiters, which are Jupiter-size planets with orbital periods of $P_{\rm orb}<10$\,days and equilibrium temperatures of $T_{\rm eq}\gtrsim1\,000$\,K. Inflated hot Jupiters are ideal targets for atmospheric studies, particularly in transmission, because of their large radii and thus atmospheric scale heights, which produce relatively high atmospheric transmission-signal amplitudes $\Delta$~\citep{sing2018}, namely,%
\begin{equation}
    \Delta=2\cdot\frac{R_{\rm p}\cdot H_{\rm s}}{R^2_{\star}}\,,
    \label{eq_transmission_signal_amplitude}
\end{equation}
where $R_{\rm p}$ is the planetary radius, $H_{\rm s}$ is atmospheric scale height, and $R_{\star}$ is the stellar radius. As a reference, for HD\,209458\,b, a typical hot-Jupiter, we have $\Delta\sim160$\,ppm (assuming an atmospheric mean molecular weight of $\mu=2.3$\,g\,mol$^{-1}$, representative of \ch{H}/\ch{He}-dominated atmospheres; \citealt{lecavelier2008}).

High-resolution spectroscopy (HRS) from ground-based facilities (see, e.g.~\citealt{birkby2018} for a review), have proven to be a valid technique to probe atmospheric features of warm- and hot-giant planets, because the large spectral resolving power ($R\gtrsim$ 25\,000) allows us to identify the characteristic spectral features of the different chemical components of an exo-atmosphere and to detect their Doppler shift due, for instance, to atmospheric dynamical effects (e.g.~\citealt{brogi2016}).

Hot (and ultra-hot) Jupiters are the most studied targets with high-resolution spectroscopy. In recent years, the sample of hot-giant planets whose atmospheres have been investigated has increased and we are entering an era of comparative exoplanet atmospheric studies (e.g.~\citealt{gandhi2023}). In this framework, increasing the statistical sample of hot Jupiters with atmospheric characterisations is important for improving our understanding of the main physical and chemical mechanisms driving their atmospheres and their formation (e.g.~\citealt{dawson2018,madhusudhan2019}).

In this work, we search for the atmospheric signature of the two hot-Jupiter planets KELT-8\,b~\citep{fulton2015} and KELT-23\,Ab~\citep{johns2019}, which both orbit Sun-like stars. Both targets were discovered using data from the Kilodegree Extremely Little Telescope survey (KELT, ~\citealt{pepper2007}) and confirmed with photometric and spectroscopic follow-up observations. As the name suggests, KELT-23\,Ab orbits around a star that is a member of a wide binary system, composed of KELT-23\,A (aka BD+66\,911\,A; aka 2MASS\,J15283520+6621314) and KELT-23\,B (aka BD+66\,911\,B; aka 2MASS\,J15283577+6621288), with a minimum mutual distance between the components of $570$\,au~\citep{johns2019}. In Table~\ref{tab_parameters}, we report a summary of the physical and orbital parameters of the two planetary systems.

Both the targets are inflated versions of Jupiter; indeed, their low density (less than half Jupiter’s) reflects the fact that they have a radius larger than Jupiter ($\sim1.6\,R_{\rm J}$ and $\sim1.3\,R_{\rm J}$, for KELT-8\,b and KELT-23\,Ab, respectively) and a slightly smaller mass ($\sim0.83\,M_{\rm J}$ and $\sim0.94\,M_{\rm J}$, for KELT-8\,b and KELT-23\,Ab, respectively). The two hot atmospheres have a large atmospheric scale height ($H_{\rm s} = 708$\,km for KELT-8\,b and $H_{\rm s}=406$\,km for KELT-23\,Ab, assuming $\mu=2.3$\,g\,mol$^{-1}$), which makes the two hot Jupiters ideal targets for atmospheric investigation. Even though KELT-23\,Ab has an $H_{\rm s}$ that is smaller than that of KELT-8\,b, the host-star has a smaller radius too, resulting in a similar atmospheric transmission signal amplitude ($\Delta=167$\,ppm for KELT-8\,b and $\Delta=156$\,ppm for KELT-23\,Ab). Although these targets are promising candidates for atmospheric investigations, there are no atmospheric studies in the literature for any of them.

The aim of this work is to search for the atmospheric signal of KELT-8\,b and KELT-23\,Ab in transmission and perform a first investigation of the chemical-physical properties of their atmospheres under simple assumptions. This paper is organised as follows. In Sect.~\ref{methods}, we describe the observations, the data reduction process, and the analyses we performed for the atmospheric characterisation of the targets. In Sect.~\ref{results}, we report the results of our atmospheric studies, while in Sect.~\ref{discussion}, we discuss the results. Finally, in Sect.~\ref{conclusion}, we report our conclusions and future perspectives.

\begin{table*}
\caption{Main physical and orbital parameters of the KELT-8 and KELT-23\,A systems.}
\label{tab_parameters}
\centering
\resizebox{0.70\textwidth}{!}{ 
\begin{tabular}{l l l c}
\hline
\hline \\[-8pt]
Parameter\tablefootmark{a} & KELT-8 & KELT-23\,A & References\tablefootmark{b}\\
\hline \\[-6pt]
\multicolumn{1}{l}{\textbf{Stellar Parameters}} \\ [2pt] %
   Spectral\ Class\dotfill & G2\,V & G2\,V & 1, 2\\
    $B_{\rm T}$\dotfill& $11.713\pm0.057$ & $11.029\pm0.049$ & 3, 3\\
    $V_{\rm T}$\dotfill& $10.925\pm0.048$ & $10.376\pm0.039$ & 3, 3\\
    $J$\dotfill& $9.586\pm0.026$ & $9.208\pm0.032$ & 4, 4\\
    $H$\dotfill& $9.269\pm0.032$ & $\geq8.951$ & 4, 4\\
    $K$\dotfill& $9.177\pm0.021$ & $\geq8.904$ & 4, 4\\
    $W1$\dotfill& $8.981\pm0.021$ & $8.754\pm0.022$ & 5, 5\\
    $W2$\dotfill& $8.983\pm0.019$ & $8.789\pm0.020$ & 5, 5\\
    $W3$\dotfill& $8.928\pm0.028$ & $8.766\pm0.022$ & 5, 5\\
    $W4$\dotfill& $8.533\pm0.339$ & $8.564\pm0.202$ & 5, 5\\   
    $M_{\star}\ $[M$_\odot$]\dotfill& $1.131^{+0.084}_{-0.079}$ & $0.944^{+0.060}_{-0.054}$ & 6, 2\\
    $R_{\star}\ $[R$_\odot$]\dotfill& $1.391\pm0.025$ & $0.996\pm0.015$ & 6, 2\\
    $T_{\textrm{eff}}$ [K]\dotfill& $5\,742\pm44$ & $5\,899\pm49$ & 6, 2\\
    $ $[Fe/H] [dex]\dotfill& $0.26\pm0.10$ & $-0.105^{+0.078}_{-0.077}$ & 6, 2\\
    log$_{10}$(g) [log$_{10}$, cgs]\dotfill& $4.19\pm0.08$ & $4.417^{+0.026}_{-0.025}$ & 6, 2\\
    $V_{\textrm{sys}}$ [km\,s$^{-1}$]\dotfill& $-34.51\pm0.20$ & $-14.99\pm0.30$ & 7, 7\\
    $ $Parallax [mas]\dotfill& $5.074\pm0.011$ & $7.871\pm0.013$ & 7, 7\\
    $ $Distance [pc]\dotfill& $197.10^{+0.42}_{-0.44}$ & $127.04\pm0.21$ & 7, 7\\
    $ $Age [Gyr]\dotfill& $6.4^{+2.9}_{-2.0}$ & $6.3^{+3.5}_{-3.2}$ & 6, 2\\[4pt]
\multicolumn{1}{l}{\textbf{Planetary Parameters}} \\ [2pt] %
    $M_\textrm{p}\ $[M$_{\rm J}$]\dotfill& $0.83\pm0.12$ & $0.938^{+0.048}_{-0.044}$ & 6, 2\\
    $R_\textrm{p}\ $[R$_{\rm J}$]\dotfill& $1.586\pm0.046$ & $1.323\pm0.025$ & 6, 2\\
    $M_\textrm{p}\ $[M$_{\oplus}$]\dotfill& $264\pm38$ & $298^{+15}_{-14}$ & 6, 2\\
    $R_\textrm{p}\ $[R$_{\oplus}$]\dotfill& $17.40\pm0.50$ & $14.83\pm0.28$ & 6, 2\\
    $\rho_{\rm p}\ $[g\,cm$^{-3}$]\dotfill& $0.276\pm0.047$ & $0.503^{+0.039}_{-0.036}$ & 6, 2\\
    $T_\textrm{eq}$ [K]\dotfill& $1\,675^{+61}_{-55}$ & $1\,561\pm20$ & 1, 2\\ [2pt]
    $P_{\rm orb}$ [days]\dotfill& $3.24408156\pm0.0000010$ & $2.25528745\pm0.00000018$ & 8, 8\\ [2pt]
    $T_\textrm{0}$ [BJD$_{\textrm{TDB}}$] \dotfill & $2\,457\,986.46737\pm0.00026$ & $2\,458\,918.461247\pm0.000021$ & 8, 8\\ [2pt]
    $i$ [deg]\dotfill& $82.65^{+0.81}_{-1.00}$ & $85.37^{+0.31}_{-0.30}$ & 1, 2\\ [2pt]
    $e$ \dotfill& $\leq0.047$ & $\leq0.05$ & 9, 2\\  [2pt]
    $a$ [au]\dotfill& $0.0382\pm0.0025$ & $0.03302^{+0.00068}_{-0.00064}$ & 6, 2\\
    $K_\textrm{p}$ [km\,s$^{-1}$]\dotfill& $149\pm26$ & $159\pm14$ & 6, 2\\  [2pt]
\hline                                            
\end{tabular}
}
\tablefoot{
        \tablefoottext{a}{The symbols of the parameters listed in the table have the following meanings: $B_{\rm T}$, $V_{\rm T}$ - apparent magnitudes in the Tycho photometric bands; $J$, $H$, $K$ - apparent magnitudes in the 2MASS photometric bands; $W1$, $W2$, $W3$, $W4$ - apparent magnitudes in the WISE photometric bands; $M_\star$ - stellar mass; $R_\star$ - stellar radius; $T_{\rm eff}$ - stellar effective temperature; [Fe/H] - stellar iron abundance; log$_{10}$(g) - logarithm of surface stellar gravity; $V_{\rm sys}$ - systemic radial velocity; $M_{\rm p}$ - planetary mass; $R_{\rm p}$ - planetary radius; $\rho_{\rm p}$ - planetary mean density; $T_{\rm eq}$ - planetary equilibrium temperature; $P_{\rm orb}$ - orbital period; $T_0$ - transit epoch; $i$ - orbital inclination; $e$ - orbital eccentricity; $a$ - orbital semi-major axis; $K_{\rm p}$ - planetary radial-velocity semi-amplitude.}\\
        \tablefoottext{b}{The left reference numbers are for the parameters of the KELT-8 system, and the right ones for those of the KELT-23\,A system. The references of the values in the table are: 1.\,\citet{fulton2015}; 2.\,\citet{johns2019}; 3.\,\citet{hog2000}; 4.\,\citet{cutri2003}; 5.\,\citet{cutri2013}; 6.\,This work and Baratella et al. (in prep.); 7.\,\citet{gaiaDR3_2023}; 8.\,\citet{kokori2023};  9.\,\citet{bonomo2017}.}
}
\end{table*}

\section{Methods}
\label{methods}
\subsection{Observations and data reduction}
\label{methods_obs}
We simultaneously collected the data sample with both GIANO-B (wavelength range: $950-2\,450$\,nm, $R\approx 50\,000$) and HARPS-N (wavelength range: $383-693$\,nm, $R\approx 115\,000$) high-resolution spectrographs, in the GIARPS configuration \citep{claudi2017}, at the Telescopio Nazionale Galileo (TNG), as part of the Global Architecture of Planetary Systems (GAPS) Project\footnote{\url{https://theglobalarchitectureofplanetarysystems.wordpress.com/}}. 
We used the HARPS-N spectra extracted with the DRSv3.7 online pipeline~\citep{pepe2002} to redetermine the stellar atmospheric and physical parameters of KELT-8 (Sect.~\ref{methods_stellar}). 
For the characterisation of the atmospheres of KELT-8\,b and KELT-23\,Ab, we only used the near-infrared (NIR) GIANO-B data, as we were primarily concerned with estimating the volatile content of the atmospheres via molecular signatures. 

The observations were taken with the nodding ABAB acquisition mode, where the target and sky spectra were taken in pairs, while alternating between two nodding positions along the slit (A and B) separated by $5^{\prime \prime}$, allowing an optimal subtraction of the detector noise and background.

For KELT-8\,b, two transit observations were available (i.e. 23 June 2020 and 19 July 2020). However, during the second transit, the weather conditions were not optimal due to the presence of calima (i.e. a meteorological phenomenon that occurs when fine sand and dust particles from the Sahara desert are lifted into the atmosphere and transported by winds), which lowered the signal-to-noise ratio (S/N). Moreover, GIANO-B showed greater instability than during the first night. For these reasons, we chose to include only the first data set in this analysis. On 23 June 2020, 70 spectra were recorded (with an exposure time $t_{\rm exp} = 200$\,s) and the target was observed at a mean airmass of $1.03$.

For KELT-23\,Ab, only one complete transit observation was available (i.e. 15 April 2023). A second transit was observed on 24 April 2023, but it was only partial due to an instrumental problem. Therefore, we did not include it in our analysis. On 15 April 2023, 60 spectra (with $t_{\rm exp} = 200$\,s) were recorded and the target was observed at a mean airmass of $1.27$. We refer to Table~\ref{logobs} for a schematic log of the observations used for our analysis.

The raw spectra were extracted and calibrated in wavelength using the GOFIO pipeline Python-3 version ~\citep{rainer2018}. The wavelength calibration performed by GOFIO is based on observing the spectrum of a U-Ne lamp as a template. Since the lamp is only observed at the end of the night to avoid persistence on the camera, the GOFIO wavelength solution is not sufficiently accurate; thus, it is expected to shift and jitter during the observations due to the mechanical instability of the instrument. To correct for the spectral drift of the instrument, we aligned all the spectra of a single night to a common reference frame via a cross-correlation with a time-averaged observed spectrum of the target used as a template. Thanks to this correction, we achieved a residual shift of the spectra well below $0.3$\,km\,s$^{-1}$ (approximately one-tenth of a pixel) for most of the spectral orders. Even though the spectrum of the calibrator was not taken simultaneously with the target spectra, we used the stationarity of telluric lines to perform an accurate wavelength calibration. In particular, after the alignment of the spectra to a common reference frame, we refined the wavelength calibration by matching a set of telluric lines in the time-averaged observed spectrum with a high-resolution atmospheric transmission model of the Earth generated via the ESO Sky Model Calculator ~\citep{noll2012} and solving for the pixel-wavelength relation with a fourth-order polynomial fit.

The GIANO-B spectral range covers the $Y,J,H,K$ bands in $50$ spectral orders (where the order of 0 is the reddest and 49 is the bluest). For our analysis, we excluded a fixed set of spectral orders dominated by tellurics (8-11, 23-26, 34-37) and the $Y$ band (orders 38-49) in which GIANO-B has a characteristic drop of throughput. Finally, for each observing night, we excluded a few orders with high residual drift, or for which the refined wavelength calibration procedure failed (i.e. orders number 3, 4, and 30 for KELT-8\,b spectra and 4, 13, 15, 18, and 30-32 for KELT-23\,Ab spectra). In this way, we were able to use a spectral range that is as more uniform as possible among the nights, where the small differences are mainly due to the stability of the pixel-wavelength solution (which, for a few orders, depends on the S/N of the observations).

\begin{table*}[!h]
\caption{GIANO-B observations of KELT-8\,b and KELT-23\,Ab.\tablefootmark{a}} 
\label{logobs}
\centering
\resizebox{0.75\textwidth}{!}{ 
\begin{tabular}{l c c c c c c}
\hline
\hline\\[-9pt]
    Target & Night & Airmass & $N_{\rm obs}$ & $t_{\rm exp}$\,[s] & (S/N)$_{\rm avg}$ & (S/N)$_{\rm min}$ to (S/N)$_{\rm max}$ \\
\hline\\[-8pt]
    KELT-8\,b & 23 June 2020 & $1.12\rightarrow1.00\rightarrow1.06$ & $70$ & $200$ & $30$ & $4-51$ \\
    KELT-23\,Ab & 15 April 2023& $1.31\rightarrow1.26\rightarrow1.27$ & $60$ & $200$ & $22$ & $5-36$ \\
\hline                                            
\end{tabular}
}
\tablefoot{
    \tablefoottext{a}{From left to right we report: the date at the start of the observing night; the airmass during the planetary transit; the number of observed spectra $N_{\rm obs}$; the exposure time per spectrum $t_{\rm exp}$; the S/N  averaged across the whole spectral range (S/N)$_{\rm avg}$; the range of S/N values (S/N)$_{\rm min}$ to (S/N)$_{\rm max}$ in the individual spectral orders.}
}
\end{table*}

\subsection{Telluric and stellar spectra removal procedure}
\label{methods_tellrem}
To remove any telluric and stellar contamination, we employed a principal component analysis (PCA), which is often used in HRS analyses (e.g.~\citealt{dekok2013}) and successfully applied to GIANO-B data in previous works (e.g.~\citealt{giacobbe2021,carleo2022,basilicata2024}).
\begin{figure*}[h]
\centering
   \includegraphics[width=15cm]{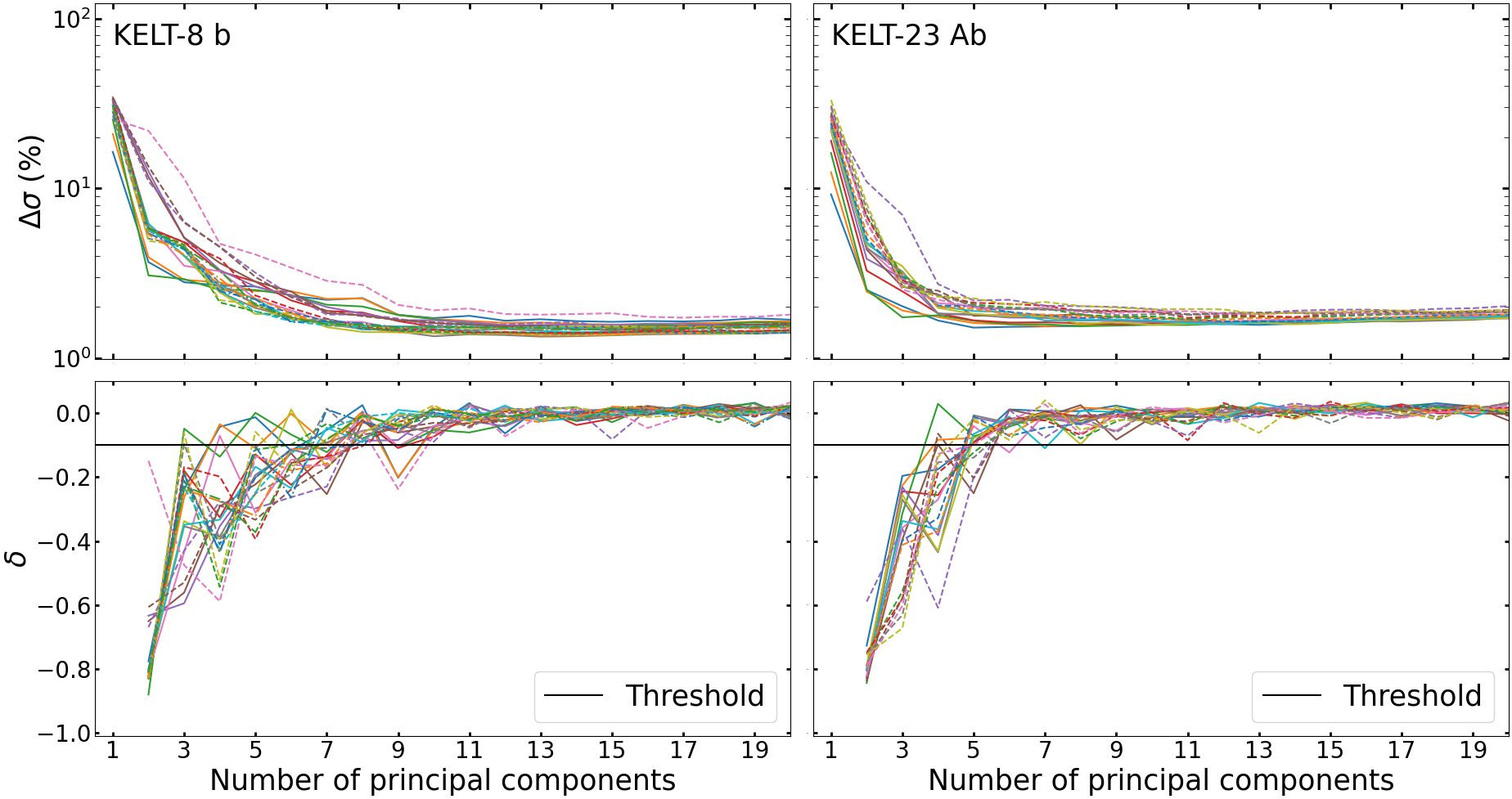}
     \caption{Variation of the standard deviation ($\sigma$) of the residuals as a function of the number of principal components removed via PCA for the KELT-8\,b (left panels) and the KELT-23\,Ab (right panels) dataset. Top panels: Relative variation $\Delta\sigma$ of the standard deviation of the residuals (expressed in percentage value and reported in logarithmic scale for better visualisation) in the different spectral orders (coloured solid and dashed lines). Bottom panels: Relative variation of $\Delta\sigma\, (\delta)$ in the different spectral orders (coloured solid and dashed lines). The solid black horizontal line represents the $-0.1$ threshold on $\delta$ used to identify the minimum number of principal components to select in order to remove the telluric and stellar contamination.}
     \label{tell_rem}
\end{figure*}
The aim of PCA is to identify common trends between different wavelength channels in the spectra as a function of time (represented by the stationary telluric lines or quasi-stationary stellar lines) to correct them. To do this, the eigenvectors of the data covariance matrix (i.e. the principal components) were computed and ordered by decreasing contribution to the global variance (identified by the associated eigenvalue). The PCA was applied to each spectral order, considered as an $M\times N$ data matrix (with $M = 60-70$ spectra and $N = 2048$ pixels). Before computing the principal components, we applied the following standardisation for each spectral order. First, we normalised each spectrum to its median value to correct the baseline flux variations. Then, we masked the spectral channels with stronger or saturated telluric lines. To do this, we subtracted the time-averaged spectrum to all the spectra (i.e. we subtracted to each spectral channel its mean value), we computed the standard deviation of data in each spectral channel and its median value ($\sigma_{\rm m}$), and then we masked the spectral channels that had a standard deviation $\sigma > 1.5 \, \sigma_{\rm m}$. After subtracting the time-averaged spectrum from all the spectra and masking highly contaminated spectral channels, we reduced the data to a null-mean variable by subtracting the mean value computed on the different spectral channels from each spectrum. The principal components were computed using the \textsc{PYDL.PCOMP}\footnote{\url{https://pydl.readthedocs.io/en/latest/api/pydl.pcomp.html}} routine. Once the principal components were calculated, we built the matrix that should mainly describe the telluric and stellar contaminations via a linear combination of the principal components and removed it from the original data matrix, obtaining a residual matrix. Finally, we applied a high-pass filter to each row of the residual matrix to remove any possible residual correlation between different spectral channels. The high-pass filtering was conducted with a boxcar average, using the \textsc{SMOOTH} IDL function~\footnote{\url{https://www.nv5geospatialsoftware.com/docs/SMOOTH.html}}, with smoothing width of $w=40$\,pixels.

To choose the appropriate number of principal components to remove, we followed the method suggested by Spring \& Birkby (in review). In particular, for each spectral order, we computed the relative variation of the standard deviation ($\sigma$) of the residuals by gradually increasing the number of components removed from the data: $\Delta\sigma=\frac{\sigma_{i-1}-\sigma_{i}}{\sigma_{i-1}}$, with $i$ = the $i$-th number of principal components removed. As seen from Fig.~\ref{tell_rem}, the $\Delta\sigma$ value of the residuals gradually decreases when the number of principal components removed increases, down to a plateau, as expected. To find the minimum number of components at which the standard deviation of the residuals does not change in a significative way (i.e. the plateau is reached), we studied the relative variation of $\Delta\sigma$ (indicated with the symbol $\delta$) to see where this variation approaches zero. We placed a threshold at $-0.1$ and selected the minimum number of principal components for which $-0.1 < \delta < 0$ for all the spectral orders (see Fig.~\ref{tell_rem}). We chose this arbitrary threshold empirically since we visually noticed that, at the plateau, the variations of $\delta$ were below the $0.1$ level. A posteriori, since we do not measure any telluric residuals in the rest of the analysis, while preserving the planetary signal, this value seems to be a sufficiently good threshold for identifying the telluric contamination, at least in our dataset. Following this procedure, the number of principal components we decided to remove for the two datasets is: $10$ for the 23 June 2020 dataset and $9$ for the 15 April 2023 dataset.
\begin{figure*}
\centering
\includegraphics[width=9cm]{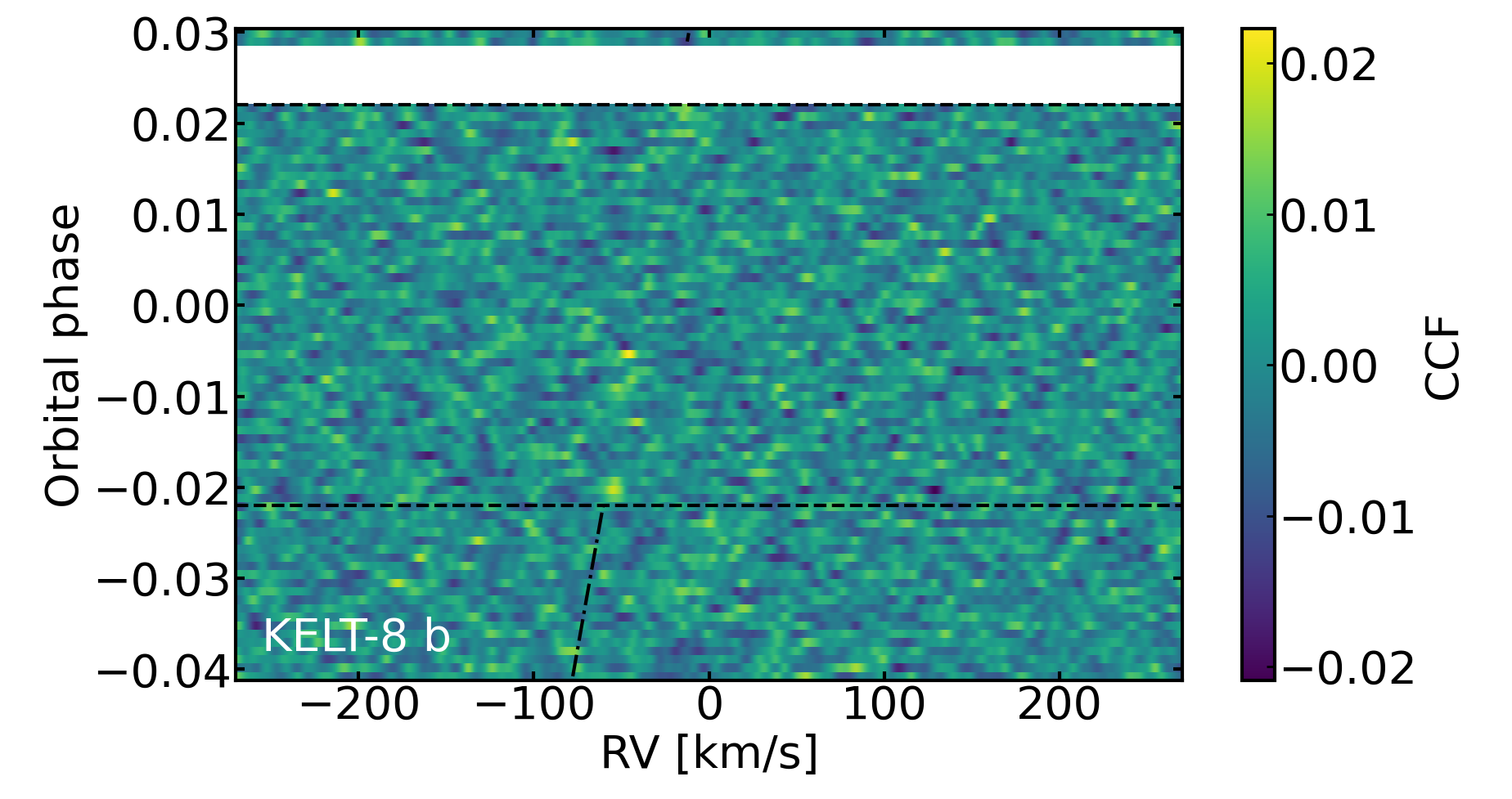}
\includegraphics[width=9cm]{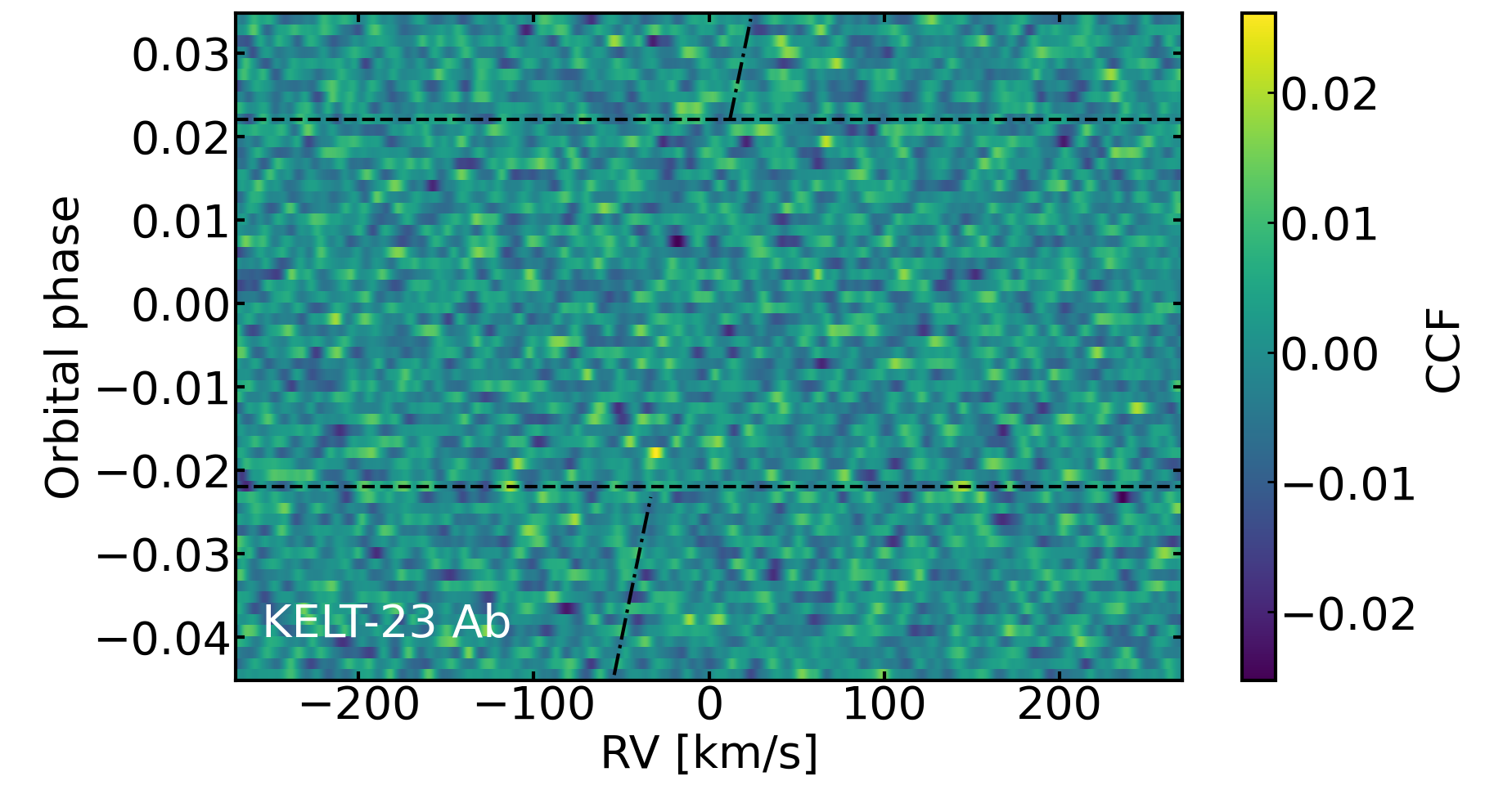}
\caption{CCF values as a function of the orbital phase computed by cross-correlating the model containing only \ch{H2O} with the data of 23 June 2020 (left panel) and 15 April 2023 (right panel). The horizontal dashed lines represent the transit ingress and egress while the dash-dotted line represents the expected CCF peak trail due to the planetary motion, as measured in the observer rest frame. The expected CCF peak trail in transit is not represented for clarity. As shown, no telluric residuals are visible by eye at ${\rm RV}=0$\,km\,s$^{-1}$ in both datasets. The white horizontal band in the left panel corresponds to a lack of out-of-transit spectra.}
\label{telluric_ccf}
\end{figure*}

Once we applied the telluric correction, we visually inspected the map of cross-correlation function (CCF) values as a function of the orbital phase obtained with the model of water that we used for the CCF analysis (see Sect.~\ref{methods_ccf}), as a further check. As we can see from the maps in Fig.~\ref{telluric_ccf}, there are no visible telluric residuals at a radial-velocity RV $=0$\,km\,s$^{-1}$. It is worth noting that due to the combined effect of Earth's velocity around the barycenter of the Solar System (barycentric velocity) and the velocity of the centre of mass of the star-planet systems with respect to the Earth (systemic velocity), the planetary signal of KELT-8\,b never overlapped with the telluric signal in terms of RV during the observation. Indeed, the signal of KELT-8\,b had a strictly negative RV (it increased from $-60$\,km\,s$^{-1}$ to $-19$\,km\,s$^{-1}$) during the planetary transit. In the case of KELT-23\,Ab, the RV of the planetary signal increased from $-32$\,km\,s$^{-1}$ to $11$\,km\,s$^{-1}$ during the transit, crossing the telluric signal at orbital phases $0.007<\phi<0.013$ (corresponding to $5$ spectra), where the planetary signal had an $|$RV$|<3$\,km\,s$^{-1}$, measured in the telluric rest frame (observer) . Since the planetary signals were shifted from the telluric rest-frame in terms of RV (of $\sim-40$\,km\,s$^{-1}$ for KELT-8\,b and $\sim-11$\,km\,s$^{-1}$ for KELT-23\,Ab, at the mid-transit point) during most of the transit, they were less affected by the telluric contamination.

\subsection{Update of the KELT-8 system parameters}
\label{methods_stellar}
Since the search for and interpretation of atmospheric signals (Sect.~\ref{methods_ccf} and Sect.~\ref{methods_retrieval}) depends on the system (stellar and planetary) parameters, we first compared the parameters of both host stars KELT-8 and KELT-23\,A given in the discovery papers \citep{fulton2015, johns2019} with those reported by \textit{Gaia} DR3 \citep{gaiaDR3_2023}. 
Noticing the discrepancy between the mass and radius of KELT-8 in \citet[$M_{\star}=1.211^{+0.078}_{-0.066}~\rm M_\odot$ and $R_{\star}=1.67^{+0.14}_{-0.12}~\rm R_\odot$]{fulton2015} and those provided by \textit{Gaia} ($M_{\star}=0.93\pm0.04~\rm M_\odot$ and $R_{\star}=1.37\pm0.03~\rm R_\odot$), we decided to recompute them.

We first redetermined the stellar atmospheric parameters (effective temperature, $T_{\rm eff}$, surface gravity, $\log g$, and iron abundance, [Fe/H]) of KELT-8 through the same procedure as other works of the GAPS Program (see \citealt{biazzo2022} and references therein). Specifically, we considered the co-added spectrum of the target built from the collected individual HARPS-N spectra and used the \textit{qoyllur-quipu} (q2\footnote{\url{https://github.com/astroChasqui/q2}}) tool developed by \cite{Ramirez2014}. This code allows for the iterative use of the 2019 MOOG version \citep{moog} to derive the stellar parameters and to measure the elemental abundances through the standard equivalent width (EW) method (i.e. $T_{\rm eff}$ and $\log g$ obtained by imposing the excitation and ionisation equilibria of \ch{FeI} and \ch{FeII} lines). The EWs of the iron transitions taken from the line list published in \citet{biazzo2022} were measured using the software ARES v2 \citep{sousa2015}, which automatically fits a Gaussian profile to the observed line profile. However, lines with EW $>120$\,m$\AA$ were manually checked with the task \textit{splot} of IRAF\footnote{IRAF is distributed by the National Optical Astronomy Observatories, which are operated by the Association of Universities for Research in Astronomy, Inc., under the cooperative agreement with the National Science Foundation. NOAO stopped supporting IRAF, see \url{https://iraf-community.github.io}.} since these are better reproduced with a Voigt profile. We then excluded from the analysis those lines with uncertainties larger than 10\%. Regarding the model atmospheres, we used the plane-parallel models linearly interpolated from the ATLAS9 grids of \cite{castelli_kurukz_2003}, computed with a solar-scale chemical composition and new opacities (ODFNEW). The final stellar atmospheric parameters ($T_{\mathrm{eff}}$, $\log g$, and [Fe/H]\footnote{[Fe/H]=$\log$(Fe)$_{\star}$-$\log$(Fe)$_{\odot}$, where $\log$(Fe)$_{\odot}$ is the solar abundance value taken from Table 2 in \cite{biazzo2022}}) are reported in Table \ref{tab_parameters}.

We determined the mass, radius, and age of the host star by modelling the stellar spectral energy distribution (SED; sampled with the magnitudes given in Table~\ref{tab_parameters}) and using the MESA Isochrones and Stellar Tracks \citep{2015ApJS..220...15P}. Specifically, we computed their best values and $1\sigma$ uncertainties from the medians and $68.3\%$ confidence intervals of the posterior distributions as obtained with the \texttt{EXOFASTv2} tool in a Bayesian differential evolution Markov chain Monte Carlo (DE-MCMC) framework \citep{2017ascl.soft10003E, 2019arXiv190709480E}. For this purpose, we adopted Gaussian priors on the spectroscopically determined $T_{\rm eff}$ and [Fe/H] as well as on the \textit{Gaia} DR3 parallax \citep{gaiaDR3_2023}. We find that the derived stellar radius is compatible with that provided by \textit{Gaia} within $1\sigma$, while the stellar mass is more in agreement with the results repored by \citet{fulton2015}. We were then able to redetermine the parameters of the hot Jupiter KELT-8\,b by using the updated stellar parameters and the transit and radial velocity parameters reported in \citet{fulton2015}, as given in our Table~\ref{tab_parameters}.

\subsection{Search for atmospheric signals through cross-correlation analysis}
\label{methods_ccf}
The high-resolution transmission spectra of the exoplanetary atmospheres are very dispersed and the S/N values for the single lines is low ($\rm{S/N}_{\rm line} \lesssim 1$). However, since thousands of spectral lines are observed simultaneously (especially in the case of the large spectral range of GIANO-B), their signals can be combined resulting in a boost in terms of S/N. Indeed, if $N$ lines of equal depth were co-added, the S/N would rise by a factor of $\sqrt{N}$~\citep{birkby2018}. The information contained in such a large number of lines can be combined by cross-correlating the residual data with template transmission spectra of the planet’s atmosphere.

To probe the presence of an atmosphere around the two hot Jupiters, we first searched for the presence of water vapour (\ch{H2O}), the most common primary molecular species in hot-giant planets' atmosphere across a broad range of atmospheric temperatures \citep{madhusudhan2012}. We built a model for each planet using the petitRADTRANS code~\citep{molliere2019}, assuming an isothermal atmosphere at the equilibrium temperature reported in Table~\ref{tab_parameters}, computed between $10$ and $10^{-8}$\,bar in pressure. The model assumes constant-with-altitude abundance (volume mixing ratio, VMR) profiles for molecular hydrogen (VMR$_{\ch{H2}} = 0.855$), helium (VMR$_{\ch{He}}=0.145$), and water (VMR$_{\ch{H2O}} = 10^{-3}$). Although single-species models do not match any specific chemical scenarios, this was the simplest framework we could adopt to probe the presence of \ch{H2O} and, thus, of trace species in addition to \ch{H2} and \ch{He}. After detecting the atmospheric \ch{H2O} signal, we also tried to search for the signal of other secondary chemical species, including \ch{CO}, \ch{CH4}, \ch{HCN}, \ch{C2H2}, \ch{CO2}, and \ch{NH3}, building single-species models following the method described earlier in this work. We note that while \ch{H2O} is expected to provide most of the opacity (and therefore searching for water alone should still result in a single-species model with meaningful continuum), a single-species model with minor species alone will not have the right continuum; thus, these models are not informative when looking to infer the bulk composition of the planet. Overall, they can only be used in this context to search for the presence of a species.

To search for the atmospheric signature, for each planet, we computed the CCF between the models and the data. The CCF was evaluated by shifting the model in wavelength on a fixed grid of RV lags:
\begin{equation}
    \Delta \rm{RV}=c\cdot\frac{\Delta\lambda}{\lambda}\,,
    \label{eq_rvlag}
\end{equation}
where $c$ is the speed of light and $\lambda$ is the wavelength of any spectral line. We explored a range of RV lags from $-270$\,km\,s$^{-1}$ to $+270$\,km\,s$^{-1}$, in steps of $0.1$\,km\,s$^{-1}$. The numeric computation was performed using the \textsc{C\_CORRELATE Pxy(L)\,IDL} function\footnote{\url{https://www.nv5geospatialsoftware.com/docs/C_CORRELATE.html}}, with null lag ($L=0$), since the RV lags were applied to the wavelength array associated with the model. For each lag, the model was interpolated (via spline interpolation) on the same wavelength array of data, before computing the CCF. The CCFs calculated for each spectral order were co-added to obtain a single CCF for each exposure of each night.

High-resolution measurements of exo-atmospheres are sensitive to the Doppler motion of the atmospheric spectral lines. In particular, if the atmospheric signal matches the cross-correlated model, a peak in the CCF is visible, and it moves in RV according to the motion-induced Doppler shift. In absence of atmospheric dynamical effects, this time-dependent Doppler shift is made up of three RV components, namely,
\begin{equation}
    V_{\rm RV}(t)=V_{\rm p}(t)+V_{\rm sys}-V_{\rm bary}(t)\,,
    \label{eqtotalrv}
\end{equation}
where $V_{\rm p}(t)$ is the planet's time-dependent RV, $V_{\rm sys}$ is the systemic velocity, and $V_{\rm bary}(t)$ is the time-dependent Earth barycentric velocity. Since the orbital eccentricity of the two planets is negligible (i.e. compatible with $0$ at $<3\,\sigma$,~\citealt{fulton2015,bonomo2017,johns2019}), we assumed a circular orbit for both targets. Therefore, the $V_{\rm p}(t)$ term can be expressed as
\begin{equation}
    V_{\rm p}(t) = K_{\rm p}\cdot \sin\left(2\pi\phi(t)\right)\,,
    \label{eqplanetrv}
\end{equation}
where $\phi(t)$ is the time-dependent planetary orbital phase and $K_{\rm p}$ is the planetary radial-velocity semi-amplitude. The latter, in case of a circular orbit, can be expressed as a function of the orbital period, $P_{\rm orb}$, orbital inclination, $i$, planetary mass, $M_{\rm p}$, and stellar mass, $M_{\star}$ only, that is,%
\begin{equation}
    K_{\rm p}= \frac{M_{\star}\sin{i}}{(M_{\rm p}+M_{\rm \star})^{\frac{2}{3}}}\cdot\left(\frac{2\pi G}{P_{\rm orb}}\right)^\frac{1}{3}\,,
    \label{eqkp}
\end{equation}
where $G$ is the gravitational constant.

If we shift into the planetary rest frame, the measured RV of the atmospheric spectrum ($V_{\rm rest}$) is $0$\,km\,s$^{-1}$ in the absence of a Doppler shift induced by atmospheric dynamics. Therefore, after having subtracted the barycentric, systemic, and planetary RV from the CCF trails (i.e. after having moved to the exoplanet rest-frame), all the CCF peaks should align at $V_{\rm rest}=0$\,km\,s$^{-1}$.

By subtracting different orbital solutions from the CCF trail (in our analysis we explored different $K_{\rm p}$), a different alignment of the CCF peaks as a function of the orbital phase is obtained. Summing in phase all the CCF values for each trial $K_{\rm p}$, the planetary signal as a function of the rest-frame velocity, $V_{\rm rest}$, is maximised and it is possible to build the so-called $K_{\rm p}-V_{\rm rest}$ maps. In these maps, in the case of the detection of an atmospheric signature, a strong peak of the signal at $V_{\rm rest} = 0$\,km\,s$^{-1}$ and the predicted $K_{\rm p}$ can be observed. From Table~\ref{tab_parameters}, we have that the predicted values of $K_{\rm p}$ are $K_{\rm p}=149\pm26$\,km\,s$^{-1}$ for KELT-8\,b and $K_{\rm p}=159\pm14$\,km\,s$^{-1}$ for KELT-23\,Ab. It is worth noting that the uncertainty on the value of $K_{\rm p}$ is mainly driven by the uncertainty on the planetary mass. Indeed, if the value of $M_{\rm p}$ for KELT-8\,b had a relative error of the order of that of KELT-23\,Ab (i.e. $5$\%), then the uncertainty on the $K_{\rm p}$ value would have been similar to that of KELT-23\,Ab (i.e. $\pm 16$\,km\,s$^{-1}$). In this work, we explore a range of $K_{\rm p}$ that spans from $0$ to $270$\,km\,s$^{-1}$ in steps of $1.0$\,km\,s$^{-1}$.

We took advantage of the high sampling of the CCF (larger than the GIANO-B pixel scale of $\sim3$\,km\,s$^{-1}$) for a precise shift of the CCF trails into the planetary rest frame. However, to avoid the use of correlated data points in our analysis, we binned the CCF values in RV using a bin width of $3.1$\,km\,s$^{-1}$. We took the median of the CCF values in each RV bin as the value of the CCF associated with each bin, before co-adding the CCF values in phase for each trial $K_{\rm p}$.

\subsection{Temperature and abundance retrieval analysis}
\label{methods_retrieval}
The CCF analysis provides us with a catalogue of the molecular species present in the atmosphere. However, it cannot determine the abundance of the species found nor the physical characteristics (e.g. temperature) of the atmospheric layers that most contribute to the observed transmission spectrum. To investigate the chemical and physical properties of the atmosphere of KELT-8\,b and KELT-23\,Ab, we need an atmospheric retrieval within a Bayesian framework.

As is fairly common in the literature (e.g.~\citealt{line2021,kasper2021,kasper2023,boucher2023}), our framework is based on the cross-correlation-to-log-likelihood remapping~\citep{brogiline2019,gibson2020}. In particular, we employed the likelihood equation described in Sect. 3.3 of~\citet{gibson2020}, but we decided not to fit the two parameters $\alpha$ (the scaling factor of the model) and $\beta$ (a scaling term for the white noise). We did not fit $\alpha$ because the scale height of the atmosphere is included in our parametrisation of the atmospheric model; therefore, there are no physical reasons to rescale the model. We did not fit $\beta$ because we directly estimated the error of each spectral channel from the data by computing the standard deviation of the time sequence of each spectral channel. In this way, the error already includes the possible presence of uncorrelated (e.g. astrophysical) jitter. Both our approach and the one by~\citet{gibson2020} rely on the same assumption, namely the absence of correlated noise in the residual spectra.

To model the exoplanet atmospheres we assumed a one-dimensional (1D) isothermal atmosphere (with temperature $T_0$) in hydrostatic equilibrium. The planetary radius corresponding to the atmospheric pressure of 0.1 bar was set equal to the value shown in Table~\ref{tab_parameters} and not retrieved. This is a common approximation for high-resolution retrievals (e.g.~\citealt{boucher2023}), in which any telluric removal procedure normalises spectra to their continuum levels, that can be avoided when low-resolution measurements are available.
We also considered a grey cloud layer by parameterising the pressure ($P_{\rm c}$) above which the atmosphere becomes completely opaque.

We modelled the chemical composition following two distinct approaches: an equilibrium chemistry approach and a `free-chemistry' approach. In the first one, chemical equilibrium is assumed and the VMR of each element is derived by the chemical network \textit{Chemcat}\footnote{\url{https://chemcat.readthedocs.io/en/main/chemistry_tutorial.html}} (Cubillos et al. in prep.) and varies accordingly with two parameters: the metallicity ([M/H]) referred to the solar value and the carbon-to-oxygen (C/O) ratio. For each trial [M/H], the C/O ratio is varied by changing only the carbon abundance, which is scaled relatively to the oxygen abundance. In the `free-chemistry' approach, the VMR of each molecule corresponds to one parameter and can vary freely.

\begin{figure}[h!]
\centering
\includegraphics[width=9cm]{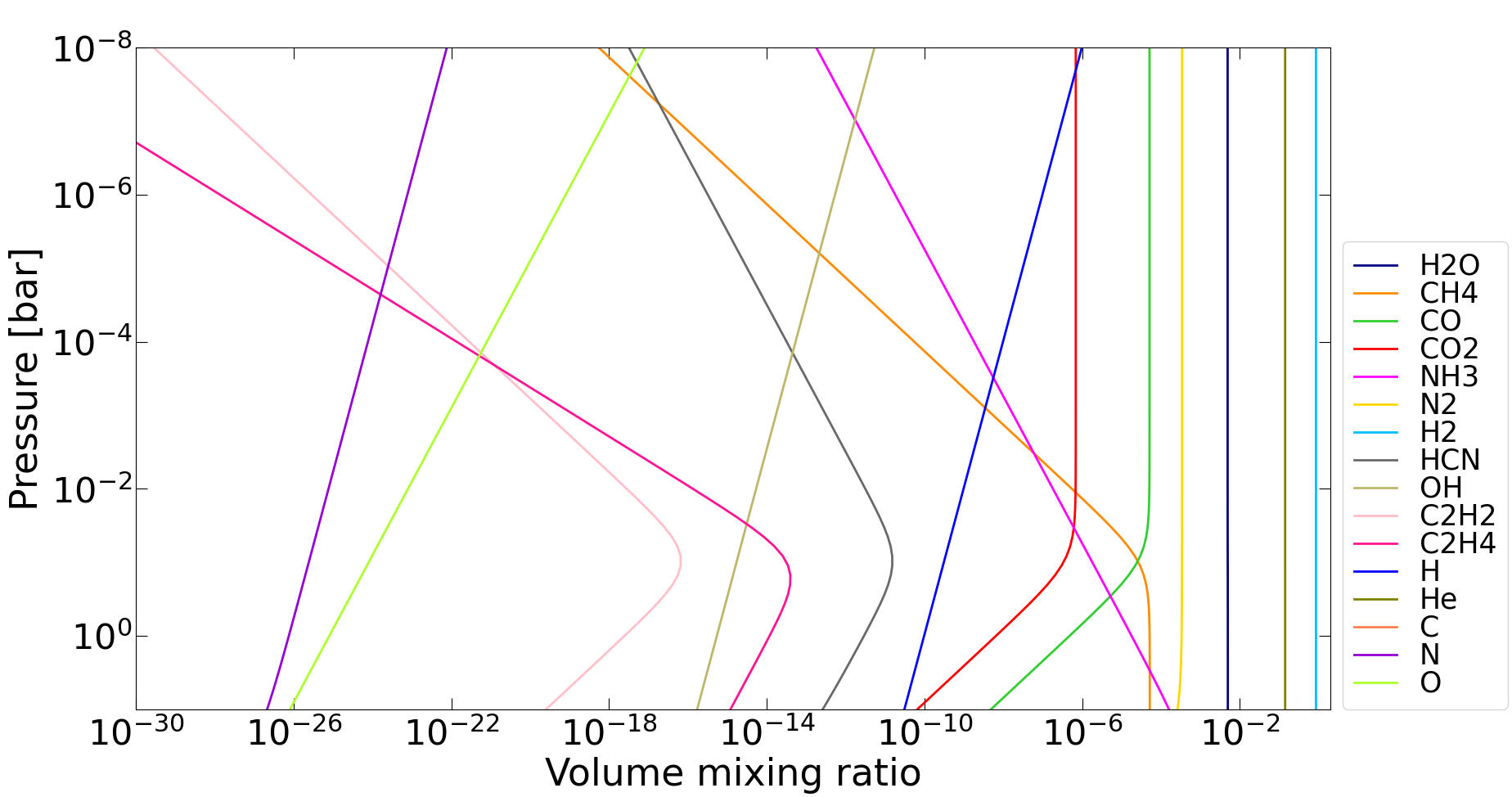}
    \caption{VMRs of different chemical species as a function of atmospheric pressure. The VMRs are computed assuming an isothermal atmosphere in chemical equilibrium with the best-fit parameters retrieved for KELT-8\,b and reported in Sect.~\ref{retrkelt8}.}
    \label{vmr_chemeq_plot}
\end{figure}

We stress that while in `free-chemistry' the VMR profiles are assumed constant-with-altitude, in chemical equilibrium they depend on the temperature and pressure profile and even an isothermal profile can lead to non-constant VMRs, as shown in Fig.~\ref{vmr_chemeq_plot}.

Although only \ch{H2O} is detected in cross-correlation for both planets, we considered for our retrievals \ch{H2O}, \ch{CO}, \ch{CH4}, \ch{CO2}, \ch{C2H2}, \ch{HCN} and \ch{NH3}.
In `free chemistry', the VMR of \ch{H2} and \ch{He} is derived assuming a constant $\ch{H2} / \ch{He}$ equal to 5.897~\citep{Asplund2009}. This means that an atmosphere of only hydrogen and helium will contain 85\% \ch{H2} and 15\% \ch{He}. The mean molecular weight is computed accordingly.

For the radiative transfer computation, we employed the petitRADTRANS code on a pressure grid of 50 layers equidistant in log-space ($-8 \leq \log_{10}[P \, (\text{bar})] \leq 1$).
The cross-sections employed in this work are summarised in Table~\ref{tab_crossections}. We also included the collision-induced absorption (CIA) cross-sections for the \ch{H2}-\ch{H2}~\citep{borysow2001,borysow2002} and \ch{H2}-\ch{He}~\citep{borysow1988,borysow1989a,borysow1989b} pairs.
\begin{table}[!h]
\caption{Cross-sections employed by petitRADTRANS.} 
\label{tab_crossections}
\centering
\resizebox{0.45\textwidth}{!}{
\begin{tabular}{l l l}
\hline
\hline
    Molecule&Cross-section&Reference\\
\hline
    \ch{H2O} & POKAZATEL &~\citet{h2o_pokazatel}\\
    \ch{CH4} & HITEMP &~\citet{ch4_hitemp}\\
    \ch{NH3} & CoYuTe &~\citet{nh3_exomol}\\
    \ch{HCN} & ExoMol &~\citet{hcn_exomol}\\
    \ch{C2H2} & aCeTY &~\citet{c2h2_acety}\\
    \ch{CO2} & Ames &~\citet{co2_ames}\\
    \ch{CO} & HITEMP &~\citet{co_hitemp}\\
\hline                                            
\end{tabular}
}
\end{table}

The transmission spectrum thus obtained had then to be convolved with the instrumental profile of GIANO-B (assumed as a Gaussian with an FWHM of $3.1$\,km\,s$^{-1}$) and shifted according to Eq.(\ref{eqtotalrv}).

When using the likelihood approach, it is essential to replicate the effects of telluric removal analysis on each model generated by the parameter sampling algorithm. This helps to prevent or mitigate potential biases caused by signal distortion induced by the filtering procedures. In the field, this analysis step is called `model reprocessing'. Apart from the name, there is no agreement in the literature on the optimal approach to model reprocessing~\citep{pino2022,brogi2023}. In this work, we followed the same recipe described in \citet{giacobbe2021} where the fitted telluric spectrum, stored after processing the observations, is multiplied by each transmission model generated by the sampler and passed through the same procedures as described above, including the PCA. 

The framework described above is implemented within a custom parallelised version of the DE-MCMC~\citep{terbraak2006,eastman2013,bonomo2015}. We employed a number of walkers $n_{\rm w} = n_{\rm f}\cdot2$, with $n_{\rm f}$ being the number of free parameters. The number of burn-in iterations discarded is variable and is computed as the minimum number of iterations for which all the chains have a likelihood value greater than the median value. The maximum number of steps is fixed to $100\,000$ but the DE-MCMC is halted when the distributions of all the parameters converge according to the Gelman-Rubin statistics~\citep{gelman1992}, following the prescription of~\citep{ford2006}.
The retrieved parameters and their uniform priors are summarised in Tables~\ref{table_chemeq} and~\ref{table_freechem}.

\section{Results}
\label{results}
\subsection{Detection of atmospheric signals}
\label{results_ccf}
\begin{figure*}[h!]
\centering
\includegraphics[width=9cm]{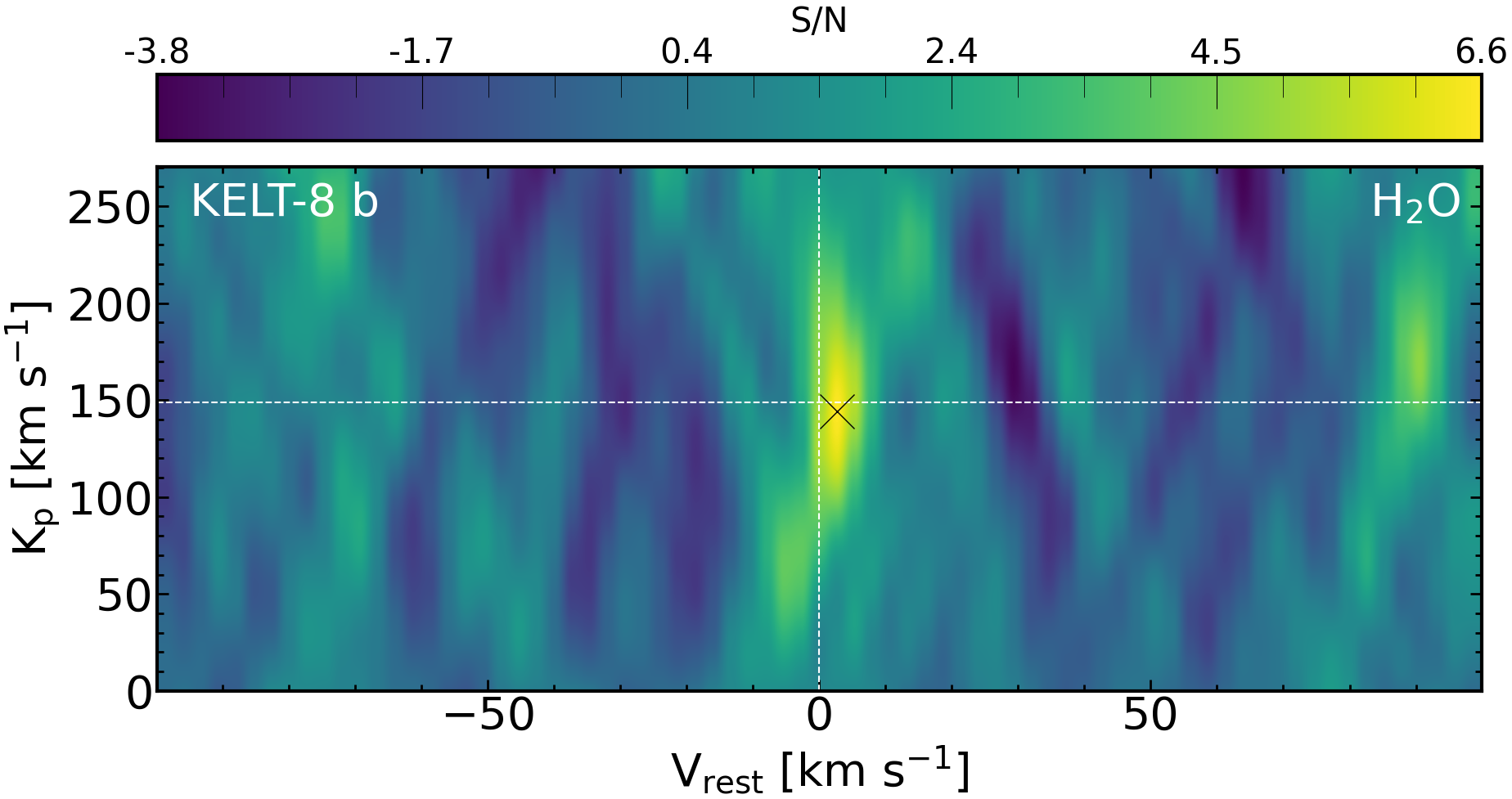}
\includegraphics[width=9cm]{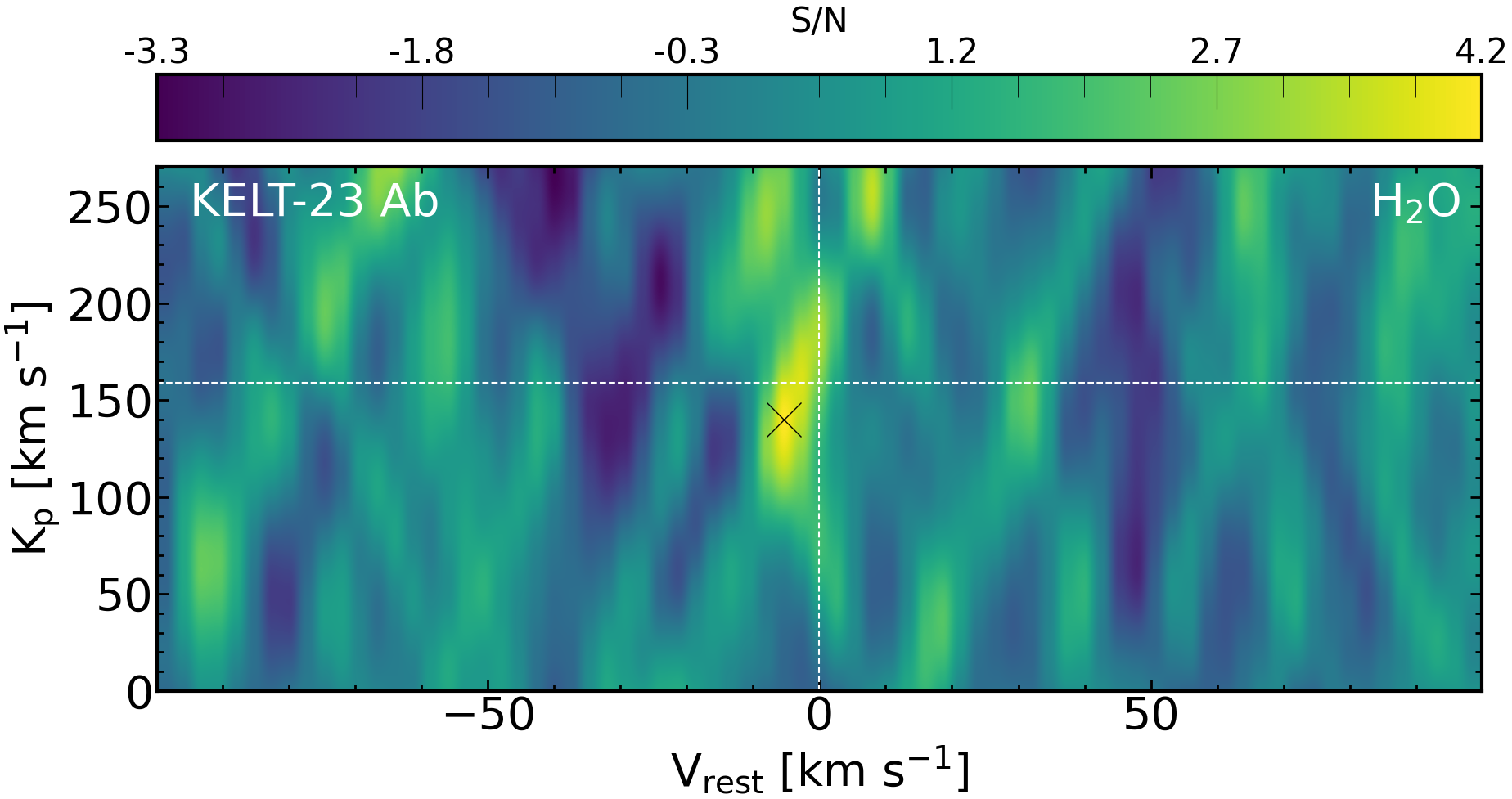}
\caption{(S/N) $K_{\rm p}-V_{\rm rest}$ maps. These maps are built by cross-correlating a single-species model of \ch{H2O} with data of KELT-8\,b (left panel) and KELT-23\,Ab (right panel), as described in the text. The maps are computed dividing the CCF values by the standard deviation of the noise far ($|V_{\rm rest}|\geq25$\,km\,s$^{-1}$) from the peak. The horizontal (vertical) white dashed lines represent the expected $K_{\rm p}$ ($V_{\rm rest}$) of the atmospheric signal. The black `$\times$' marks denote the cross-correlation maximum in each plot.}
\label{kpvrest_map}
\end{figure*}
We built the S/N $K_{\rm p}-V_{\rm rest}$ maps by converting the CCF values into detection S/N. To do so, we divided the CCF values by the standard deviation of the noise far ($|V_{\rm rest}|\geq25$\,km\,s$^{-1}$) from the peak. In Fig.~\ref{kpvrest_map}, we report the (S/N) $K_{\rm p}-V_{\rm rest}$ maps for the \ch{H2O} signal, where we can see that we detect the atmospheric signals of the two targets. In particular, we detect the presence of \ch{H2O} with an S/N of $6.6$ and $4.2$ in the atmosphere of KELT-8\,b and KELT-23\,Ab, respectively. In the case of KELT-8\,b, the peak of the signal is at a radial-velocity semi-amplitude $K_{\rm p}=144^{+33}_{-32}$\,km\,s$^{-1}$ (the uncertainty on this value is computed assuming as the extremes of the $1\,\sigma$ interval the $K_{\rm p}$ values where the S/N drops by $1$\,unity from the peak), compatible with the expected one at $<1\,\sigma$ and at a $V_{\rm rest}=2.7^{+2.7}_{-2.7}$\,km\,s$^{-1}$. In the case of KELT-23\,Ab, the peak is shifted at $K_{\rm p}=140^{+26}_{-29}$\,km\,s$^{-1}$ ($\sim20$\,km\,s$^{-1}$ smaller than the expected value but compatible at $1\,\sigma$) and at a $V_{\rm rest}=-5.4^{+2.7}_{-2.7}$\,km\,s$^{-1}$. The origins of these RV shifts are discussed in Sect.~\ref{discussion_atmo}. Thus, our detection of the \ch{H2O} signal constitutes the first measurement of the atmospheric signal of the two targets. For what concerns the other chemical species, we did not find any significant CCF signal (S/N$\geq3$) for either planet. Even though we do not detect the other secondary chemical species, we cannot exclude their presence in the atmosphere of the two targets, as discussed in Sect.~\ref{discussion}.

\subsection{Atmospheric-retrieval analyses}
\label{results_retrieval}
In this section, we report the first investigation of the atmospheric properties of the two targets. As described in Sect.~\ref{methods_retrieval}, we performed an atmospheric retrieval with the simplest assumption of an isothermal atmosphere in chemical equilibrium and a second retrieval in `free-chemistry'. We report the retrieved value of the fitted parameters and the associated uncertainties in Table~\ref{table_chemeq} and Table~\ref{table_freechem} for the chemical-equilibrium and `free-chemistry' retrieval, respectively. We took the medians and the $15.87\%$ and $84.14\%$ quantiles of the posterior distributions as the values and $1\,\sigma$ uncertainties of the fitted parameters.

\begin{table}[!h]
\caption{Parameters, best-fitting values, and adopted uniform ($U$) priors for the chemical-equilibrium retrieval.} 
\label{table_chemeq}
\resizebox{\hsize}{!}{
\centering
\begin{tabular}{l c c l}
\hline
\hline
    Parameter&KELT-8\,b\tablefootmark{a}&KELT-23\,Ab\tablefootmark{a}&Prior\\
\hline\\[-0.5pt]
    $K_{\rm p}$ [km\,s$^{-1}$]&$154\pm10$&$163^{+25}_{-26}$&$U$[0; 250]\\ [3pt]
    $V_{\rm rest}$ [km\,s$^{-1}$]&$3.30^{+0.89}_{-0.76}$&$-3.14^{+2.50}_{-1.93}$&$U$[-10; 10]\\ [3pt]
    log$_{10}(P_{\rm c})$ [log$_{10}$, bar]&$\geq-2.29$&$\geq-4.58$&$U$[-8; 2]\\ [3pt]
    $ $[M/H] [dex]&$0.77^{+0.61}_{-0.89}$&$-0.42^{+1.56}_{-1.35}$&$U$[-3; 3]\\ [3pt]
    log$_{10}$(C/O)&$\leq-0.52$&$\leq-0.11$&$U$[-3; 2]\\ [3pt]
    $T_{\rm 0}$ [K]&$897^{+162}_{-113}$&$1833^{+612}_{-667}$&$U$[300; 3000]\\ [3pt]
\hline                                            
\end{tabular}
}
\tablefoot{
    \tablefoottext{a}{For each target, we report the best-fitted value and the $1\,\sigma$ confidence interval of the retrieved parameters with the exception of log$_{10}$($P_{\rm c}$) and log$_{10}$(C/O) for which we report the $2\,\sigma$-level lower and upper limit, respectively.}
}
\end{table}

\begin{table}[!h]
\caption{Parameters, best-fitting values, and adopted uniform ($U$) priors for the `free-chemistry' retrieval.} 
\label{table_freechem}
\resizebox{\hsize}{!}{
\centering
\begin{tabular}{l c c l}
\hline
\hline
    Parameter&KELT-8\,b\tablefootmark{a}&KELT-23\,Ab\tablefootmark{a}&Prior\\
\hline\\[-0.5pt]
    $K_{\rm p}$ [km\,s$^{-1}$]&$152^{+11}_{-12}$&$152^{+21}_{-22}$&$U$[0; 250]\\ [3pt]
    $V_{\rm rest}$ [km\,s$^{-1}$]&$3.21^{+0.96}_{-1.02}$&$-3.88^{+1.86}_{-1.40}$&$U$[-10; 10]\\ [3pt]
    log$_{10}(P_{\rm c})$ [log$_{10}$, bar]&$\geq-2.50$&$\geq-4.88$&$U$[-8; 2]\\ [3pt]
    $T_{\rm 0}$ [K]&$880^{+198}_{-148}$&$2306^{+490}_{-765}$&$U$[300; 3000]\\ [3pt]
    log$_{10}($VMR$_{\rm H_2O})$ & $-2.07^{+0.53}_{-0.72}$ & $-2.26^{+0.75}_{-1.24}$ &$U$[-10; -1]\\ [3pt]
    log$_{10}($VMR$_{\rm CO})$ & $\leq-2.55$ & $\leq-2.14$ & $U$[-10; -1]\\ [3pt]
    log$_{10}($VMR$_{\rm CH_4})$ & $\leq-3.52$ & $\leq-2.17$ & $U$[-10; -1]\\ [3pt]
    log$_{10}($VMR$_{\rm CO_2})$ & $\leq-2.49$ & $\leq-2.16$ & $U$[-10; -1]\\ [3pt]
    log$_{10}($VMR$_{\rm C_2H_2})$ & $\leq-2.14$ & $\leq-1.87$ & $U$[-10; -1]\\ [3pt]
    log$_{10}($VMR$_{\rm HCN})$ & $\leq-4.40$ & $\leq-2.25$ & $U$[-10; -1]\\ [3pt]
    log$_{10}($VMR$_{\rm NH_3})$ & $\leq-4.47$ & $\leq-3.38$ & $U$[-10; -1]\\ [3pt]
\hline                                            
\end{tabular}
}
\tablefoot{
    \tablefoottext{a}{For each target, we report the best-fitted value and the $1\,\sigma$ confidence interval of the retrieved parameters. The values of the log$_{10}$($P_{\rm c}$) and the VMRs reported for all the chemical species but \ch{H2O} are the $2\,\sigma$-level lower and upper limits, respectively.}
}
\end{table}
\subsubsection{KELT-8\,b}
\label{retrkelt8}
\begin{figure*}[h!]
\centering
\includegraphics[width=14cm]{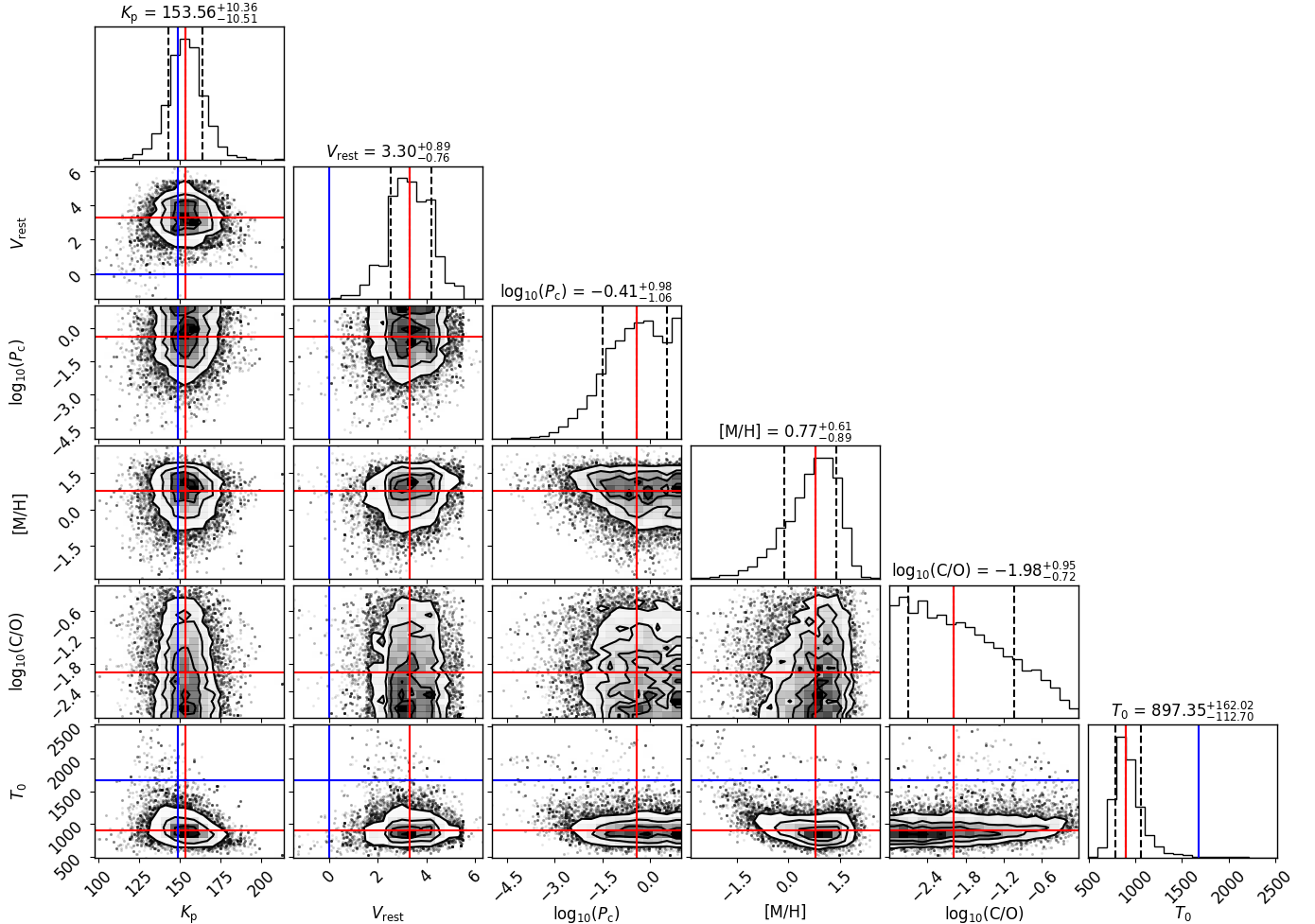}
    \caption{Posterior distributions for all the parameters of the chemical-equilibrium retrieval, for KELT-8\,b. Off-diagonal plots report the 2D posterior distribution for pairs of parameters with the $1\,\sigma$, $2\,\sigma$, and $3\,\sigma$ confidence intervals. On-diagonal plots report the posterior distributions of each parameter marginalised over the remaining parameters. The blue lines represent the predicted values for some parameters and the red lines represent the median of the posterior distributions.}
    \label{cornerkelt8}
\end{figure*}
In Fig.~\ref{cornerkelt8}, we report the corner plots with retrieved posterior distributions of the fitted atmospheric parameters for the chemical-equilibrium retrieval for KELT-8\,b. For this target, the atmospheric retrieval in chemical equilibrium converges at the expected planetary orbital solution ($K_{\rm p}$) at a confidence level $<1\,\sigma$. The retrieval converges to a $V_{\rm rest}=3.30^{+0.89}_{-0.76}$\,km\,s$^{-1}$, confirming the residual redshift of the CCF peak visible in the left panel of Fig.~\ref{kpvrest_map}. We find a lower limit on the cloud-top pressure of $P_{\rm c}\geq5.13$\,mbar (at $2\,\sigma$ level). For KELT-8\,b, we are able to constrain a metallicity of [M/H] $=0.77^{+0.61}_{-0.89}$\,dex, which corresponds to $3.2$ times the stellar metallicity (i.e. [Fe/H] $=0.26\pm0.10$\,dex). With respect to the C/O ratio, we are only able to place an upper limit; in particular, we retrieve a sub-solar C/O ratio (i.e. C/O $\leq0.30$, at $2\,\sigma$ level). The non-constrained log$_{10}$(C/O) distribution, with a median at C/O$=0.01$, is due to the lack of detections for any carbon-bearing species with our dataset. A sub-unity C/O ratio was expected given these non-detections and the simultaneous detection of \ch{H2O}, the most common oxygen-bearing species. Even if we put an upper limit on the C/O ratio, the posterior probability distribution seems to increase when the C/O ratio decreases (at the lower extreme of the $1\,\sigma$ interval we have C/O $=0.002$). New observations that would allow us to detect the presence of secondary carbon-bearing species could help improve the constraints on the C/O ratio posterior probability distribution.

With respect to the temperature, the retrieval converges at $T_{0}=897^{+162}_{-113}$\,K. This value is lower than the estimated equilibrium temperature (see Table~\ref{tab_parameters}). It is worth noting that with transmission spectroscopy, it is possible to probe only a small fraction of the atmosphere (i.e. the terminator) between the two hemispheres of planets. Since hot Jupiters are expected to be tidally locked due to their proximity to the host stars, the average temperatures of the two faces of these planets can be very different ($\Delta T\sim100-1\,000$\,K,~\citealt{heng2015}). As a consequence, the estimated equilibrium temperature could not be representative of the average temperature at the terminator, which is strongly dependent on the atmospheric circulation. In addition, the retrieved isothermal temperature is only an ``indicative'' temperature of the atmospheric layers we manage to probe, where most of the absorption occurs. Thus, it is reasonable to get temperatures lower than the equilibrium one. Indeed, studying the transmission contribution function of the water spectrum for this planet, we find that the core of the lines of \ch{H2O} is produced in the layers with pressure $10^{-5} < P < 10^{-4}$\,bar, where the temperature is expected to be lower than the equilibrium one (see e.g. the temperature-pressure profile reported in the bottom-middle panel of `Extended Data' Fig.~2 in~\citealt{giacobbe2021}, for a typical hot Jupiter).\\
It is worth noting that even assuming that the equilibrium temperature is representative of the average temperature at the terminator, retrieving a temperature value lower than the expected equilibrium temperature has proven to be a common output in the literature when 1D atmospheric models are used to fit transmission spectra (e.g.~\citealt{macdonald2020} and references therein,~\citealt{boucher2023},~\citealt{xue2024}). This effect could be due to possible asymmetries in the two regions (i.e. the morning and evening one) of the terminator~\citep{macdonald2020}. However, these differences can be explored effectively only through at least a 2D atmospheric model and with a higher-quality dataset.

As can be seen in Table~\ref{table_freechem}, also the atmospheric retrieval in the `free-chemistry' scenario converges to $K_{\rm p}$ and $V_{\rm rest}$ values in accordance with those found by the chemical equilibrium retrieval at $<1\,\sigma$. Also, the retrieved value of $T_{0}$ is in accordance with the value retrieved in the former scenario at $1\,\sigma$. For \ch{H2O}, we retrieve a volume mixing ratio of log$_{10}$(VMR$_{\rm H_2O})=-2.07^{+0.53}_{-0.72}$, in accordance at the $1\,\sigma$ level with what is predicted assuming an isothermal atmosphere in chemical equilibrium at the temperature, C/O ratio, and metallicity retrieved in the chemical equilibrium scenario (i.e. log$_{10}$(VMR$_{\rm H_2O})=-2.31$); whereas for the other chemical species, we only find upper limits on the abundances, in accordance with the non-detections we find with the CCF analysis of our dataset.

\subsubsection{KELT-23\,Ab}
\label{retrkelt23}
\begin{figure*}[h!]
\centering
\includegraphics[width=14cm]{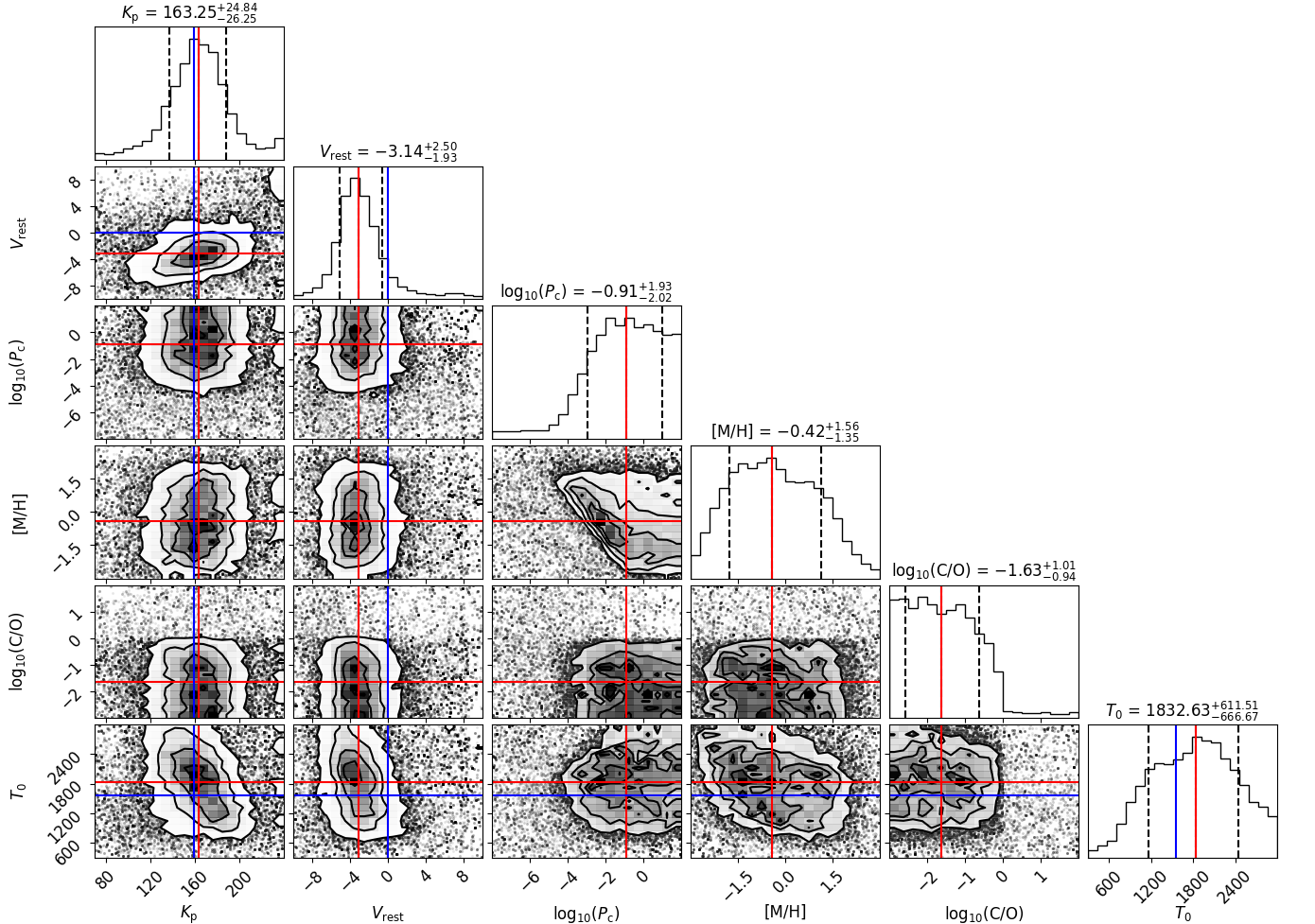}
    \caption{Posterior distributions for all the parameters of the chemical-equilibrium retrieval, for KELT-23\,Ab. Off-diagonal plots report the 2D posterior distribution for pairs of parameters with the $1\,\sigma$-, $2\,\sigma$-, and $3\,\sigma$-confidence intervals. On-diagonal plots report the posterior distributions of each parameter marginalised over the remaining parameters. The blue lines represent the predicted values for some parameters and the red lines represent the median of the posterior distributions.}
    \label{cornerkelt23}
\end{figure*}
In Fig.~\ref{cornerkelt23}, we report the corner plots with the retrieved posterior distributions of the fitted atmospheric parameters for KELT-23\,Ab under the chemical-equilibrium assumption. In the case of KELT-23\,Ab, the atmospheric retrieval converges to a value of $K_{\rm p}$ that is compatible with the expected one at $<1\,\sigma$ and to a $V_{\rm rest}$ value lower than $0$\,km\,s$^{-1}$ (i.e. $V_{\rm rest}=-3.14^{+2.50}_{-1.93}$\,km\,s$^{-1}$), confirming the residual blueshift of the signal visible in the CCF peak in the $K_{\rm p}-V_{\rm rest}$ map (Fig.~\ref{kpvrest_map}, right panel). Also for KELT-23\,Ab, we find a lower limit on the cloud-top pressure, however, in this case, it is at a lower pressure: $P_{\rm c}\geq0.026$\,mbar (at $2\,\sigma$ level). For KELT-23\,Ab, the posterior distributions of the fitted parameters are less constrained with respect to the case of KELT-8\,b, probably due to the lower S/N of the observations (as can be seen in Table~\ref{logobs}). Indeed, with respect to the chemical parameters, the metallicity posterior distribution is broader and we observe a larger degeneracy between the log$_{10}(P_{\rm c})$ and [M/H]. However, we are still able to place a first constraint on the metallicity of the atmosphere: the $1\,\sigma$ interval around the retrieved median value ranges from a sub-solar ([M/H] $=-1.77$\,dex) to a super-solar metallicity ([M/H] $=1.14$\,dex), with a median sub-solar metallicity value of [M/H] $=-0.42$\,dex, which corresponds to $0.48$ times the stellar metallicity (i.e. [Fe/H] $=-0.105^{+0.078}_{-0.077}$\,dex). For the C/O ratio, also in this case, due to the lack of detection of carbon-bearing species in our dataset, we are only able to retrieve an upper limit, in particular, we measure a sub-unity C/O ratio (i.e. C/O $\leq0.78$ at $2\,\sigma$ level). The constraint on the retrieved temperature is broad and even if the median retrieved temperature is higher than the expected equilibrium one, there is compatibility at the $1\,\sigma$ level. As can be seen in Table~\ref{table_freechem}, the atmospheric retrieval in the `free-chemistry' scenario converges to a $K_{\rm p}$ that is in accordance with the predicted value and with the value found in the chemical equilibrium scenario at $1\,\sigma$. From both atmospheric retrievals in the two analysed scenarios, we do not find any evidence for the $\sim20$\,km\,s$^{-1}$ shift in $K_{\rm p}$ visible in the $K_{\rm p}-V_{\rm rest}$ map, probably because between the peak of the CCF signal and the CCF value at the expected $K_{\rm p}$ there is a small S/N difference of $0.5$, making the peak shift statistically non-significant. However, it is interesting to note that a $K_{\rm p}$ shift of $\sim20$\,km\,$^{-1}$ is very similar to the expected shift predicted by Eq.~(7) in ~\citet{wardenier2023} due to the planetary rotation during the transit. Indeed, the predicted $K_{\rm p}$ shift is $\Delta K_{\rm p}=-21.6$\,km\,s$^{-1}$ for the equatorial rotation velocity of KELT-23\,Ab ($V_{\rm eq}=2.98$\,km\,s$^{-1}$, in case of synchronous rotation), so we cannot exclude that the observed shift is a hint of planetary rotation; however, we do find that other measurements are needed to retrieve this dynamical effect in a statistically robust way. The shift in $V_{\rm rest}$ (i.e. $V_{\rm rest}=-3.88^{+1.86}_{-1.40}$\,km\,s$^{-1}$) is in accordance at $<1\,\sigma$ with the value found assuming the chemical equilibrium and confirms a global blueshift of the atmospheric signal of KELT-23\,Ab.\\
Also, in this case, the retrieved reference temperature is in accordance with the value retrieved in the chemical equilibrium scenario at the $1\,\sigma$ level. In the case of KELT-23\,Ab, we retrieve a volume mixing ratio of \ch{H2O} equal to log$_{10}$(VMR$_{\rm H_2O})=-2.26^{+0.75}_{-1.24}$, in accordance at the $1\,\sigma$ level with what is predicted assuming an isothermal atmosphere in chemical equilibrium at the temperature, C/O ratio, and metallicity retrieved in the chemical equilibrium scenario (i.e. log$_{10}$(VMR$_{\rm H_2O})=-3.51$). In this case as well, we can only put upper limits on the abundances of other secondary chemical-species that we did not detect in cross-correlation in our dataset.

\section{Discussion}
\label{discussion}
\subsection{KELT-8\,b and KELT-23\,Ab atmospheric properties}
\label{discussion_atmo}
The two hot Jupiters analysed in this work have an atmosphere in which water vapour is the dominant secondary species. In particular, considering the $1\,\sigma$ interval of the water VMR, for KELT-8\,b we measure an abundance of water between $0.2\%$ and $3\%$, with a median value of $0.9\%$; whereas for KELT-23\,Ab, we measure an abundance of water between $0.03\%$ and $3\%$, with a median value of $0.5\%$. Even though the constraints on KELT-23\,Ab atmospheric metallicity are broader than those on KELT-8\,b, the median value of the distribution is about one order of magnitude smaller, implying an average lower abundance of volatile heavy elements (e.g. oxygen, carbon), in accordance with the lower abundance of water vapour. While we identify the \ch{H2O} signature in cross-correlation for both targets, we do not observe the presence of other secondary species in our dataset. This pushes the retrieved C/O ratios to lower values and, since we are unable to measure the abundance of carbon-bearing species, we only retrieve upper limits.

This first atmospheric analysis of KELT-8\,b and KELT-23\,Ab reveals two O-rich atmospheres, with a sub-solar C/O ratio for KELT-8\,b and a sub-unity C/O ratio for KELT-23\,Ab, at $2\,\sigma$ level. Assuming an atmosphere in chemical equilibrium at $T_{\rm eq}$ for the two planets and with C/O ratio upper limits we find, water is expected to be the most abundant secondary species in the atmosphere of hot Jupiters (e.g., \citealt{madhusudhan2012,moses2014}); whereas the formation of hydrocarbons such as \ch{C2H2} and \ch{HCN} is not favoured by the under-abundance of carbon, in accordance with our non-detections. In this scenario, the most abundant carbon-bearing species is expected to be \ch{CO}, whose formation is favoured with respect to \ch{CH4}. We find no evidence for \ch{CO} in our dataset. Since most of the \ch{CO} absorption is concentrated in two absorption bands (one centered at $1\,600$\,nm and one at $2\,400$\,nm), we also tried to probe the presence of \ch{CO} by using only GIANO-B orders in which these two bands fall (i.e. orders number 0, 1, 14, 15, 16, and 17, with the exclusion of 15 for KELT-23\,Ab) but, also in this case, we do not find any significant signal at the expected planetary RV. It is worth noting that the expected \ch{CO} abundance assuming chemical equilibrium (with the retrieved atmospheric parameters) is $\log_{10}$(VMR$_{\rm CO})=-4.3$ and $-5.1$, for KELT-8\,b and KELT-23\,Ab, respectively. These values are well below the upper limits we found from the `free-chemistry' retrieval. Indeed, the non-detection of the other probed chemical species less abundant than water does not exclude that these species are present in the atmosphere of the two targets and that their lower signal may be detectable with more observations at both high- and low-spectral resolution.

Hot Jupiters are expected to be tidally locked due to their proximity to their host stars. The strong temperature gradient between the day side and the night side of these planets can induce the presence of strong atmospheric dynamical effects (e.g super-rotating equatorial jet streams and day-to-night side winds) spreading the heat around the planets~\citep{showman2011,heng2015}. The possible presence of day-to-night side winds (with speeds of a few km\,s$^{-1}$), transporting atmospheric gases across the terminator from the sub-stellar point towards the colder side of the planet (i.e. towards the observer during the planetary transit), introduces a blueshift of the atmospheric transmission signal. Looking at the signal of \ch{H2O} for KELT-23\,Ab, we measure a residual blueshift of the signal that can be explained by the presence of a day-to-night wind on this planet with a speed of $V_{\rm wind}=-3.88^{+1.86}_{-1.40}$\,km\,s$^{-1}$ (the most precise value among the two retrieved). This value is in line with the typical wind speed measured on other hot Jupiters (e.g.~\citealt{snellen2010,brogi2016,line2021,nortmann2025}). Both targets orbit around G2\,V stars but, considering the orbits, KELT-8\,b receives $25\%$ less flux from its host star than KELT-23\,Ab. Therefore, the effects due to the redistribution of heat in the atmosphere could be less intense because the amount of stellar energy received is lower. However, this does not exclude the possible presence of large-scale circulation effects on KELT-8\,b that redistribute the incoming energy across the atmosphere and planetary interior.\\
Interestingly,~\citet{wardenier2023} investigated the effect of atmospheric `3D-ness' and the dynamics on the CCF signal of ultra-hot Jupiters, which are the most ideal targets for atmospheric investigation due to their extremely hot and bloated atmospheres.~\citet{wardenier2023} showed that complex combinations of rotation, wind patterns, 3D temperature fields and distributions of chemical species, and geometrical effects could introduce residual RV shifts of the cross-correlation signal both in $K_{\rm p}$ and $V_{\rm rest}$. Since these general results remain valid also for warmer planets, the observed residual redshift of the signal of KELT-8\,b could be due to complex atmospheric dynamical effects.\\
For example, a possible asymmetry in the intensity of the \ch{H2O} signals produced by the two limbs of the terminator can lead to this type of Doppler shift. In particular, a residual redshift of the signal could be introduced if most of the water absorption comes from the leading (morning) limb of the terminator. In this configuration, the \ch{H2O} signal is mostly produced by the portion of the atmosphere that is moving away from the observer during the transit, due to planetary rotation. Since this redshifted signal is not counterbalanced by the (reduced) blueshifted signal of the trailing (evening) limb, which is moving towards the observer, the total signal acquires a residual redshift.~\citet{wardenier2024} detected a residual redshift of the signal of \ch{H2O} for the ultra-hot Jupiter WASP-121\,b, suggesting that a possible configuration that produces the aforementioned asymmetry could be the presence of clouds on the trailing limb of the planet but not on the leading limb. The presence of `evening clouds' would mute the signal of \ch{H2O} only on the side of the terminator moving towards the observer. However, as~\citet{wardenier2024} pointed out, in this scenario, we would simultaneously have to explain why clouds do not prevail on the leading limb of the planet; for example by assuming condensation of \ch{H2O} at the trailing limb and gravitational settlement in the night side of the planet or cloud-patchiness. Another possible configuration could be a lack of water in the trailing limb of the terminator due to \ch{H2O} dissociation on the dayside of the planet. This was proposed as a possible explanation by~\citet{mansfield2024}, for instance, who detected a residual redshift of the \ch{H2O} signal for the ultra-hot Jupiter WASP-76\,b. However, the temperature of the dayside of KELT-8\,b is probably not sufficiently high to induce thermal dissociation of water (the equilibrium temperature of KELT-8\,b is $\sim500$\,K lower than the one of WASP-76\,b) and, thus, it does not seem to be a plausible configuration.\\
Due to our limited dataset, an accurate analysis of the atmospheric dynamics of KELT-8\,b (and KELT-23\,Ab) is beyond the scope of this work, as it would require more precise measurements, along with observations of the signals of other chemical species populating different atmospheric regions and their variation as a function of the orbital phase, as well as the use of more realistic atmospheric models (e.g.~\citealt{gandhi2022,beltz2023,wardenier2023,nortmann2025}). For these reasons, further studies are needed to confirm and fully understand the origin of our detection of positive $V_{\rm rest}$ for the \ch{H2O} signal of KELT-8\,b. It is worth noting that if future analyses with higher quality datasets will measure asymmetries in the two limbs of the terminator, then our retrieved values could be biased and it is reasonable to expect differences in the measured abundance values, according to~\citet{macdonald2020}. However, due to the relatively broad constraints we put on the chemical composition of the target, we expect these differences to be within the uncertainty intervals we give. This is supported by the fact that according to~\citet{macdonald2020}, these biases should be stronger for hotter planets (i.e. ultra-hot Jupiters).

Finally, it is worth noting that besides the atmospheric dynamical effects, a small orbital eccentricity can also produce the residual Doppler shift of the signal observed for both planets. Indeed, it is possible to obtain a residual radial-velocity shift of the signal between $-10$\,km\,s$^{-1}<V_{\rm rest}<10$\,km\,s$^{-1}$ by considering a small orbital eccentricity value of $e\sim0.01-0.05$ (below the upper limit reported in the literature for both planets) and different values of the planetary argument of periastron, $\omega_{\rm p}$. Thus, even though for KELT-8\,b and KELT-23\,Ab, we do not have robust evidence for an eccentric orbit in the literature, a small residual eccentricity remains a possible explanation for the observed atmospheric signal Doppler shift.

\subsection{Possible formation scenarios for the two planets}
\label{discussion_formation}
By measuring the abundances of the different chemical species populating the atmosphere of an exoplanet, it is possible to constrain its formation and evolutionary paths. Indeed, observables such as the atmospheric elemental abundance ratios (e.g. C/O and N/O) and the metallicity depend on the planet formation location into the protoplanetary disk, the accretion mechanisms that enriched the planet atmosphere with heavy elements, and the migration mechanisms that led to the current orbital configuration of the planetary system (e.g.~\citealt{mordasini2016,madhusudhan2019,pacetti2022}).

Even though we only detect the presence of \ch{H2O} in the present work, we were able to estimate its abundance and place constraints on the planetary atmospheric metallicity and upper limits on the C/O ratio for both targets. Therefore, we can try to make preliminary inferences on possible formation scenarios for the two planets. Of course, with more precise future measurements of the atmospheric composition, it would be possible to put stronger constraints on the planetary formation and evolutionary paths followed by both targets.

The observed enrichment in the two giant planets suggests the accretion of the disk gas as the source of their atmospheric metallicity~\citep[e.g.][]{turrini2021,pacetti2022}. It also suggests that the possible sequestration of \ch{O} by rock-forming elements \citep{fonte2023} should be limited, meaning that the abundance of \ch{H2O} provides a proxy for the bulk abundance of \ch{O}. This scenario would be confirmed by the non-detection of abundant rock-forming elements in future analyses. The super-solar mixing ratio estimated for \ch{H2O} in both planets argues for the accretion of high-metallicity gas enriched by the sublimation of volatile species from drifting pebbles \citep[e.g.][]{booth2019,schneider2021}. Finally, the non-detection of \ch{C} and \ch{N} species and the upper limits to the C/O ratios of both planets (see below for discussion) point to the accretion of gas from a disk region enriched primarily in \ch{O}.

Focusing on KELT-8\,b, the availability of stellar abundances from the Hypatia Catalog Database \citep{Hinkel2014} enables for a more detailed analysis. The host star exhibits a super-solar metallicity and a slightly sub-solar C/O ratio ($\sim 0.51$). Comparing these with our retrieved parameters reveals that the giant planet is enriched in heavy elements by a factor of about 3 compared to its host star, with a sub-stellar upper limit for the normalised C/O ratio ($<0.59$). This sub-stellar C/O ratio, combined with a super-stellar VMR for \ch{H2O}, suggests that the planet has accreted gas more enriched in oxygen than carbon, consistent with gas accretion from within the \ch{H2O} snowline. In particular, the Hypatia Catalog Database reports that the [O/H] of KELT-8 is 0.2 dex, meaning that it is 1.58 times higher than the solar one. The estimated VMR for \ch{H2O} then points to a $\sim7\times$ enrichment in O with respect to the star, which is compatible with the estimated super-stellar metallicity. Specifically, in a solar-like mixture, \ch{O} accounts for about 45\% of the mass of heavy elements \citep{lodders2010}, meaning that a $\sim7\times$ enrichment in \ch{O} causes a $3\times$ enrichment in metallicity, which is consistent with the estimated value.

The lack of compositional data on KELT-23\,A and the large uncertainties affecting the planetary metallicity stand in the way of us refining the discussion to the level of KELT-8\,b. Based on the correlations between stellar metallicity and composition discussed in \cite{dasilva2024}, however, the average C/O values for stars whose metallicities fall in the uncertainty range of KELT-23\,A's [Fe/H] are expected to vary between 0.45 and 0.65. While the estimated upper limit to the C/O ratio of KELT-23\,Ab does not allow us to rule out super-stellar values, it is broadly consistent with the accretion of O-enriched gas across the snowline of \ch{H2O}, as discussed in the case of KELT-8\,b. This scenario would suggest planetary metallicity values higher than the one reported in Table~\ref{table_chemeq}. Similarly to what we have discussed for KELT-8\,b, if we assume a solar-like mixture, the \ch{O} enrichment suggested by the \ch{H2O} VMR would point to a planetary metallicity [M/H] of the order of $0.5$\,dex; while this result is higher, it is still compatible within $1\,\sigma$ with the estimated atmospheric value.

Finally, it is worth noting that in order to completely understand the formation and orbital evolution path followed by KELT-23\,Ab, it is also necessary to take into consideration the presence of the outer stellar companion in a wide orbit. Indeed, under some conditions, eccentric orbital interactions with a distant companion star via Kozai–Lidov oscillations~\citep{kozai1962} can lead to a giant planet migration towards its host star and subsequent orbital circularisation through tidal friction ~\citep{fabrycky2007}. To effectively consider the magnitude of such effects in the KELT-23\,Ab case, more data about the orbital configuration of the KELT-23 system and the physical characteristics of the outer stellar body are needed.

\section{Summary and conclusions}
\label{conclusion}
In this work, we searched for the atmospheric signal of the two hot Jupiters KELT-8\,b and KELT-23\,Ab, and we started exploring their chemical and physical properties under the simplest assumption (i.e. 1D isothermal atmosphere), in order to enlarge the sample of known hot Jupiters with atmospheric detection and characterisation.

We performed our study by analysing one transit observation for each target taken with the NIR GIANO-B high-resolution ($R\approx50\,000$) spectrograph mounted at the $3.58$\,m TNG telescope. By cross-correlating the data with atmospheric transmission template models, we detect the signal of \ch{H2O} for both planets, with a CCF peak value corresponding to S/N $=6.6$ and S/N $=4.2$ for KELT-8\,b and KELT-23\,Ab, respectively, while finding no significant signal for the other probed chemical species. This constitutes the first observation of the atmospheric signal from KELT-8\,b and KELT-23\,Ab.

We explored the chemical and physical properties of the two atmospheres in a Bayesian framework. In particular, we ran two retrievals for both planets: a retrieval assuming chemical equilibrium and a `free chemistry' retrieval. Both the retrievals point towards an atmosphere rich in \ch{H2O} (from $\sim0.1\%$ to $\sim1\%$) for both targets. 

For KELT-8\,b, in the chemical-equilibrium scenario, we find an atmospheric metallicity of ${\rm [M/H]}= 0.77^{+0.61}_{-0.89}$\,dex, which corresponds to $3.2$ times the stellar metallicity, while for the C/O ratio, we can place a sub-solar upper limit at C/O $\leq0.30$ (at $2\,\sigma$ confidence level). In the `free-chemistry' scenario, we can constrain an \ch{H2O} abundance of log$_{10}$(VMR$_{\rm H_2O})=-2.07^{+0.53}_{-0.72}$, in accordance with the prediction of chemical-equilibrium, while for the other probed chemical species we put upper limits on their abundance. 

For KELT-23\,Ab, the posterior distributions of the parameters are broader than the ones retrieved for KELT-8\,b, probably due to the lower S/N of the dataset. Indeed, in the case of KELT-23\,Ab, we find an atmospheric metallicity that ranges from a sub- to super-solar metallicity in the $1\,\sigma$ confidence interval, with a sub-solar median value ([M/H] $= -0.42^{+1.56}_{-1.35}$\,dex), which corresponds to $0.48$ times the stellar metallicity. For the atmosphere of KELT-23\,Ab, we put a sub-unity C/O ratio upper limit (C/O $\leq0.78$ at $2\,\sigma$ level). The `free-chemistry' retrieval converges to an \ch{H2O} abundance of log$_{10}$(VMR$_{\rm H_2O})=-2.26^{+0.75}_{-1.24}$, in agreement with the prediction of chemical equilibrium; also in this case, we place upper limits on the abundance of the other probed chemical species.

By comparing the atmospheric chemical characteristics with those of the host stars, we obtained indications of the formation mechanisms of the two hot Jupiters. In particular, the constraints on VMR$_{\rm H_2O}$ and the upper limits on the C/O ratio suggest for both planets the accretion of O-rich gas. This, in turn, points to an accretion of gaseous material that occurred within the \ch{H2O} snowline in a pebble-rich disk, where the gas phase is enriched in oxygen due to sublimation of water ice from the inward-drifting pebbles.

With our analysis, we are able to offer a preliminary investigation of the atmospheric properties of KELT-8\,b and KELT-23\,Ab. New measurements of atmospheric spectra of both targets could allow us to detect the presence of other secondary species, in addition to refining the current \ch{H2O} abundance, thus allowing us to place stronger constraints on the metallicity and the C/O ratio posterior distributions. We stress that our main goal was not to give a fully comprehensive picture of the chemical and physical phenomena occurring in the atmosphere of the two targets, but to provide a first general characterisation of the main properties of the two atmospheres under simple assumptions. Even if more complex 2D and 3D atmospheric models (including e.g. global circulation and limb asymmetries, more parametrised temperature profiles, and different chemical scenarios) would certainly provide a more realistic description of the chemical and physical properties of the atmosphere of the two targets, this is beyond the scope of this work (also given the limited amount of data available for both targets).

Our analysis, based on only one transit observation for each target, demonstrates that these two planets, whose atmospheres were unstudied before this work, are very interesting targets for atmospheric investigations and that it is worth observing them with higher diameter instruments both at high resolution from the ground and at low resolution from space with HST and JWST.

\begin{acknowledgements}
The work is based on observations made with the Italian Telescopio Nazionale Galileo (TNG) operated on the island of La Palma by the Fundacion Galileo Galilei of the INAF (Istituto Nazionale di Astrofisica) at the Spanish Observatorio del Roque de los Muchachos of the Instituto de Astrofisica de Canarias.
The authors acknowledge financial contributions from PRIN INAF 2019 and INAF GO Large Grant 2023 GAPS-2 as well as from the European Union - Next Generation EU RRF M4C2 1.1 PRIN MUR 2022 project 2022CERJ49 (ESPLORA).
E.P. and D.T. acknowledge the support of the Italian National Institute of Astrophysics (INAF) and of the Italian Space Agency (ASI) through the ASI-INAF grant no. 2021-5-HH.0 plus addenda no. 2021-5-HH.1-2022 and 2021-5-HH.2-2024, the ASI–INAF grant no. 2016-23-H.0 plus addendum no. 2016-23-H.2-2021, and the support of the European Research Council via the Horizon 2020 Framework Programme ERC Synergy ``ECOGAL'' Project GA-855130.
L.M. acknowledges financial contribution from PRIN MUR 2022 project 2022J4H55R.
M.P. acknowledges support from the European Union – NextGenerationEU (PRIN MUR 2022 20229R43BH) and the "Programma di Ricerca Fondamentale INAF 2023".
This paper is supported by the Fondazione ICSC, Spoke 3 Astrophysics and Cosmos Observations, National Recovery and Resilience Plan (Piano Nazionale di Ripresa e Resilienza, PNRR) Project ID CN\_00000013 ``Italian Research Center on High-Performance Computing, Big Data and Quantum Computing'' funded by MUR Missione 4, Componente 2, Investimento 1.4: Potenziamento strutture di ricerca e creazione di ``campioni nazionali di R\&S (M4C2-19)'' - Next Generation EU (NGEU).
The authors acknowledge the Italian center for Astronomical Archives (IA2, https://www.ia2.inaf.it), part of the Italian National Institute for Astrophysics (INAF), for providing technical assistance, services and supporting activities of the GAPS collaboration.
This publication makes use of data products from the Wide-field Infrared Survey Explorer, which is a joint project of the University of California, Los Angeles, and the Jet Propulsion Laboratory/California Institute of Technology, funded by the National Aeronautics and Space Administration.
\end{acknowledgements}

\bibliographystyle{aa} 
\bibliography{bibl.bib}

\begin{thebibliography}{88}
\expandafter\ifx\csname natexlab\endcsname\relax\def\natexlab#1{#1}\fi

\bibitem[{{Asplund} {et~al.}(2009){Asplund}, {Grevesse}, {Sauval}, \& {Scott}}]{Asplund2009}
{Asplund}, M., {Grevesse}, N., {Sauval}, A.~J., \& {Scott}, P. 2009, \araa, 47, 481

\bibitem[{{Banzatti} {et~al.}(2020){Banzatti}, {Pascucci}, {Bosman}, {Pinilla}, {Salyk}, {Herczeg}, {Pontoppidan}, {Vazquez}, {Watkins}, {Krijt}, {Hendler}, \& {Long}}]{banzatti2020}
{Banzatti}, A., {Pascucci}, I., {Bosman}, A.~D., {et~al.} 2020, \apj, 903, 124

\bibitem[{Barber {et~al.}(2013)Barber, Strange, Hill, Polyansky, Mellau, Yurchenko, \& Tennyson}]{hcn_exomol}
Barber, R.~J., Strange, J.~K., Hill, C., {et~al.} 2013, \mnras, 437, 1828

\bibitem[{{Basilicata} {et~al.}(2024){Basilicata}, {Giacobbe}, {Bonomo}, {Scandariato}, {Brogi}, {Singh}, {Di Paola}, {Mancini}, {Sozzetti}, {Lanza}, {Cubillos}, {Damasso}, {Desidera}, {Biazzo}, {Bignamini}, {Borsa}, {Cabona}, {Carleo}, {Ghedina}, {Guilluy}, {Maggio}, {Mainella}, {Micela}, {Molinari}, {Molinaro}, {Nardiello}, {Pedani}, {Pino}, {Poretti}, {Southworth}, {Stangret}, \& {Turrini}}]{basilicata2024}
{Basilicata}, M., {Giacobbe}, P., {Bonomo}, A.~S., {et~al.} 2024, \aap, 686, A127

\bibitem[{Beltz {et~al.}(2023)Beltz, Rauscher, Kempton, Malsky, \& Savel}]{beltz2023}
Beltz, H., Rauscher, E., Kempton, E. M.-R., Malsky, I., \& Savel, A.~B. 2023, \aj, 165, 257

\bibitem[{{Biazzo} {et~al.}(2022){Biazzo}, {D'Orazi}, {Desidera}, {Turrini}, {Benatti}, {Gratton}, {Magrini}, {Sozzetti}, {Baratella}, {Bonomo}, {Borsa}, {Claudi}, {Covino}, {Damasso}, {Di Mauro}, {Lanza}, {Maggio}, {Malavolta}, {Maldonado}, {Marzari}, {Micela}, {Poretti}, {Vitello}, {Affer}, {Bignamini}, {Carleo}, {Cosentino}, {Fiorenzano}, {Giacobbe}, {Harutyunyan}, {Leto}, {Mancini}, {Molinari}, {Molinaro}, {Nardiello}, {Nascimbeni}, {Pagano}, {Pedani}, {Piotto}, {Rainer}, \& {Scandariato}}]{biazzo2022}
{Biazzo}, K., {D'Orazi}, V., {Desidera}, S., {et~al.} 2022, \aap, 664, A161

\bibitem[{{Birkby}(2018)}]{birkby2018}
{Birkby}, J.~L. 2018, arXiv e-prints, arXiv:1806.04617

\bibitem[{{Bitsch} {et~al.}(2022){Bitsch}, {Schneider}, \& {Kreidberg}}]{bitsch2022}
{Bitsch}, B., {Schneider}, A.~D., \& {Kreidberg}, L. 2022, \aap, 665, A138

\bibitem[{{Bonomo} {et~al.}(2017){Bonomo}, {Desidera}, {Benatti}, {Borsa}, {Crespi}, {Damasso}, {Lanza}, {Sozzetti}, {Lodato}, {Marzari}, {Boccato}, {Claudi}, {Cosentino}, {Covino}, {Gratton}, {Maggio}, {Micela}, {Molinari}, {Pagano}, {Piotto}, {Poretti}, {Smareglia}, {Affer}, {Biazzo}, {Bignamini}, {Esposito}, {Giacobbe}, {H{\'e}brard}, {Malavolta}, {Maldonado}, {Mancini}, {Martinez Fiorenzano}, {Masiero}, {Nascimbeni}, {Pedani}, {Rainer}, \& {Scandariato}}]{bonomo2017}
{Bonomo}, A.~S., {Desidera}, S., {Benatti}, S., {et~al.} 2017, \aap, 602, A107

\bibitem[{{Bonomo} {et~al.}(2015){Bonomo}, {Sozzetti}, {Santerne}, {Deleuil}, {Almenara}, {Bruno}, {D{\'\i}az}, {H{\'e}brard}, \& {Moutou}}]{bonomo2015}
{Bonomo}, A.~S., {Sozzetti}, A., {Santerne}, A., {et~al.} 2015, \aap, 575, A85

\bibitem[{{Booth} \& {Ilee}(2019)}]{booth2019}
{Booth}, R.~A. \& {Ilee}, J.~D. 2019, \mnras, 487, 3998

\bibitem[{{Borysow}(2002)}]{borysow2002}
{Borysow}, A. 2002, \aap, 390, 779

\bibitem[{{Borysow} \& {Frommhold}(1989)}]{borysow1989b}
{Borysow}, A. \& {Frommhold}, L. 1989, \apj, 341, 549

\bibitem[{{Borysow} {et~al.}(1989){Borysow}, {Frommhold}, \& {Moraldi}}]{borysow1989a}
{Borysow}, A., {Frommhold}, L., \& {Moraldi}, M. 1989, \apj, 336, 495

\bibitem[{{Borysow} {et~al.}(2001){Borysow}, {Jorgensen}, \& {Fu}}]{borysow2001}
{Borysow}, A., {Jorgensen}, U.~G., \& {Fu}, Y. 2001, \jqsrt, 68, 235

\bibitem[{{Borysow} {et~al.}(1988){Borysow}, {Frommhold}, \& {Birnbaum}}]{borysow1988}
{Borysow}, J., {Frommhold}, L., \& {Birnbaum}, G. 1988, \apj, 326, 509

\bibitem[{{Boucher} {et~al.}(2023){Boucher}, {Lafreni{\'e}re}, {Pelletier}, {Darveau-Bernier}, {Radica}, {Allart}, {Artigau}, {Cook}, {Debras}, {Doyon}, {Gaidos}, {Benneke}, {Cadieux}, {Carmona}, {Cloutier}, {Cort{\'e}s-Zuleta}, {Cowan}, {Delfosse}, {Donati}, {Fouqu{\'e}}, {Forveille}, {Grankin}, {H{\'e}brard}, {Martins}, {Martioli}, {Masson}, \& {Vinatier}}]{boucher2023}
{Boucher}, A., {Lafreni{\'e}re}, D., {Pelletier}, S., {et~al.} 2023, \mnras, 522, 5062

\bibitem[{{Brogi} {et~al.}(2016){Brogi}, {de Kok}, {Albrecht}, {Snellen}, {Birkby}, \& {Schwarz}}]{brogi2016}
{Brogi}, M., {de Kok}, R.~J., {Albrecht}, S., {et~al.} 2016, \apj, 817, 106

\bibitem[{Brogi {et~al.}(2023)Brogi, Emeka-Okafor, Line, Gandhi, Pino, Kempton, Rauscher, Parmentier, Bean, Mace, Cowan, Shkolnik, Wardenier, Mansfield, Welbanks, Smith, Fortney, Birkby, Zalesky, Dang, Patience, \& Désert}]{brogi2023}
Brogi, M., Emeka-Okafor, V., Line, M.~R., {et~al.} 2023, \aj, 165, 91

\bibitem[{{Brogi} \& {Line}(2019)}]{brogiline2019}
{Brogi}, M. \& {Line}, M.~R. 2019, \aj, 157, 114

\bibitem[{Carleo {et~al.}(2022)Carleo, Giacobbe, Guilluy, Cubillos, Bonomo, Sozzetti, Brogi, Gandhi, Fossati, Turrini, Biazzo, Borsa, Lanza, Malavolta, Maggio, Mancini, Micela, Pino, Poretti, Rainer, Scandariato, Schisano, Andreuzzi, Bignamini, Cosentino, Fiorenzano, Harutyunyan, Molinari, Pedani, Redfield, \& Stoev}]{carleo2022}
Carleo, I., Giacobbe, P., Guilluy, G., {et~al.} 2022, \aj, 164, 101

\bibitem[{{Castelli} \& {Kurucz}(2003)}]{castelli_kurukz_2003}
{Castelli}, F. \& {Kurucz}, R.~L. 2003, in Modelling of Stellar Atmospheres, ed. N.~{Piskunov}, W.~W. {Weiss}, \& D.~F. {Gray}, Vol. 210, A20

\bibitem[{Chubb {et~al.}(2020)Chubb, Tennyson, \& Yurchenko}]{c2h2_acety}
Chubb, K.~L., Tennyson, J., \& Yurchenko, S.~N. 2020, \mnras, 493, 1531

\bibitem[{{Claudi} {et~al.}(2017){Claudi}, {Benatti}, {Carleo}, {Ghedina}, {Guerra}, {Micela}, {Molinari}, {Oliva}, {Rainer}, {Tozzi}, {Baffa}, {Baruffolo}, {Buchschacher}, {Cecconi}, {Cosentino}, {Fantinel}, {Fini}, {Ghinassi}, {Giani}, {Gonzalez}, {Gonzalez}, {Gratton}, {Harutyunyan}, {Hernandez}, {Lodi}, {Malavolta}, {Maldonado}, {Origlia}, {Sanna}, {Sanjuan}, {Scuderi}, {Seemann}, {Sozzetti}, {Perez Ventura}, {Hernandez Diaz}, {Galli}, {Gonzalez}, {Riverol}, \& {Riverol}}]{claudi2017}
{Claudi}, R., {Benatti}, S., {Carleo}, I., {et~al.} 2017, EPJP, 132, 364

\bibitem[{Coles {et~al.}(2019)Coles, Yurchenko, \& Tennyson}]{nh3_exomol}
Coles, P.~A., Yurchenko, S.~N., \& Tennyson, J. 2019, \mnras, 490, 4638

\bibitem[{{Cutri} {et~al.}(2003){Cutri}, {Skrutskie}, {van Dyk}, {Beichman}, {Carpenter}, {Chester}, {Cambresy}, {Evans}, {Fowler}, {Gizis}, {Howard}, {Huchra}, {Jarrett}, {Kopan}, {Kirkpatrick}, {Light}, {Marsh}, {McCallon}, {Schneider}, {Stiening}, {Sykes}, {Weinberg}, {Wheaton}, {Wheelock}, \& {Zacarias}}]{cutri2003}
{Cutri}, R.~M., {Skrutskie}, M.~F., {van Dyk}, S., {et~al.} 2003, VizieR Online Data Catalog, II/246

\bibitem[{{Cutri} {et~al.}(2013){Cutri}, {Wright}, {Conrow}, {Fowler}, {Eisenhardt}, {Grillmair}, {Kirkpatrick}, {Masci}, {McCallon}, {Wheelock}, {Fajardo-Acosta}, {Yan}, {Benford}, {Harbut}, {Jarrett}, {Lake}, {Leisawitz}, {Ressler}, {Stanford}, {Tsai}, {Liu}, {Helou}, {Mainzer}, {Gettngs}, {Gonzalez}, {Hoffman}, {Marsh}, {Padgett}, {Skrutskie}, {Beck}, {Papin}, \& {Wittman}}]{cutri2013}
{Cutri}, R.~M., {Wright}, E.~L., {Conrow}, T., {et~al.} 2013, {VizieR Online Data Catalog: AllWISE Data Release (Cutri+ 2013)}, VizieR On-line Data Catalog: II/328. Originally published in: IPAC/Caltech (2013)

\bibitem[{{da Silva} {et~al.}(2024){da Silva}, {Danielski}, {Delgado Mena}, {Magrini}, {Turrini}, {Biazzo}, {Tsantaki}, {Rainer}, {Helminiak}, {Benatti}, {Adibekyan}, {Sanna}, {Sousa}, {Casali}, \& {Van der Swaelmen}}]{dasilva2024}
{da Silva}, R., {Danielski}, C., {Delgado Mena}, E., {et~al.} 2024, \aap, 688, A193

\bibitem[{{Dawson} \& {Johnson}(2018)}]{dawson2018}
{Dawson}, R.~I. \& {Johnson}, J.~A. 2018, \araa, 56, 175

\bibitem[{{de Kok} {et~al.}(2013){de Kok}, {Brogi, M.}, {Snellen, I.A.G.}, {Birkby, J.}, {Albrecht, S.}, \& {de Mooij, E.J.W.}}]{dekok2013}
{de Kok}, {Brogi, M.}, {Snellen, I.A.G.}, {et~al.} 2013, \aap, 554, A82

\bibitem[{{Eastman}(2017)}]{2017ascl.soft10003E}
{Eastman}, J. 2017, {EXOFASTv2: Generalized publication-quality exoplanet modeling code}, Astrophysics Source Code Library, record ascl:1710.003

\bibitem[{Eastman {et~al.}(2013)Eastman, Gaudi, \& Agol}]{eastman2013}
Eastman, J., Gaudi, B.~S., \& Agol, E. 2013, \pasp, 125, 83

\bibitem[{{Eastman} {et~al.}(2019){Eastman}, {Rodriguez}, {Agol}, {Stassun}, {Beatty}, {Vanderburg}, {Gaudi}, {Collins}, \& {Luger}}]{2019arXiv190709480E}
{Eastman}, J.~D., {Rodriguez}, J.~E., {Agol}, E., {et~al.} 2019, arXiv e-prints, arXiv:1907.09480

\bibitem[{{Fabrycky} \& {Tremaine}(2007)}]{fabrycky2007}
{Fabrycky}, D. \& {Tremaine}, S. 2007, \apj, 669, 1298

\bibitem[{{Fonte} {et~al.}(2023){Fonte}, {Turrini}, {Pacetti}, {Schisano}, {Molinari}, {Polychroni}, {Politi}, \& {Changeat}}]{fonte2023}
{Fonte}, S., {Turrini}, D., {Pacetti}, E., {et~al.} 2023, \mnras, 520, 4683

\bibitem[{{Ford}(2006)}]{ford2006}
{Ford}, E.~B. 2006, \apj, 642, 505

\bibitem[{{Fulton} {et~al.}(2015){Fulton}, {Collins}, {Gaudi}, {Stassun}, {Pepper}, {Beatty}, {Siverd}, {Penev}, {Howard}, {Baranec}, {Corfini}, {Eastman}, {Gregorio}, {Law}, {Lund}, {Oberst}, {Penny}, {Riddle}, {Rodriguez}, {Stevens}, {Zambelli}, {Ziegler}, {Bieryla}, {D'Ago}, {DePoy}, {Jensen}, {Kielkopf}, {Latham}, {Manner}, {Marshall}, {McLeod}, \& {Reed}}]{fulton2015}
{Fulton}, B.~J., {Collins}, K.~A., {Gaudi}, B.~S., {et~al.} 2015, \apj, 810, 30

\bibitem[{{Gaia Collaboration} {et~al.}(2023){Gaia Collaboration}, {Vallenari}, {Brown}, {Prusti}, {de Bruijne}, {Arenou}, {Babusiaux}, {Biermann}, {Creevey}, {Ducourant}, {Evans}, {Eyer}, {Guerra}, {Hutton}, {Jordi}, {Klioner}, {Lammers}, {Lindegren}, {Luri}, {Mignard}, {Panem}, {Pourbaix}, {Randich}, {Sartoretti}, {Soubiran}, {Tanga}, {Walton}, {Bailer-Jones}, {Bastian}, {Drimmel}, {Jansen}, {Katz}, {Lattanzi}, {van Leeuwen}, {Bakker}, {Cacciari}, {Casta{\~n}eda}, {De Angeli}, {Fabricius}, {Fouesneau}, {Fr{\'e}mat}, {Galluccio}, {Guerrier}, {Heiter}, {Masana}, {Messineo}, {Mowlavi}, {Nicolas}, {Nienartowicz}, {Pailler}, {Panuzzo}, {Riclet}, {Roux}, {Seabroke}, {Sordo}, {Th{\'e}venin}, {Gracia-Abril}, {Portell}, {Teyssier}, {Altmann}, {Andrae}, {Audard}, {Bellas-Velidis}, {Benson}, {Berthier}, {Blomme}, {Burgess}, {Busonero}, {Busso}, {C{\'a}novas}, {Carry}, {Cellino}, {Cheek}, {Clementini}, {Damerdji}, {Davidson}, {de Teodoro}, {Nu{\~n}ez Campos}, {Delchambre}, {Dell'Oro}, {Esquej},
  {Fern{\'a}ndez-Hern{\'a}ndez}, {Fraile}, {Garabato}, {Garc{\'\i}a-Lario}, {Gosset}, {Haigron}, {Halbwachs}, {Hambly}, {Harrison}, {Hern{\'a}ndez}, {Hestroffer}, {Hodgkin}, {Holl}, {Jan{\ss}en}, {Jevardat de Fombelle}, {Jordan}, {Krone-Martins}, {Lanzafame}, {L{\"o}ffler}, {Marchal}, {Marrese}, {Moitinho}, {Muinonen}, {Osborne}, {Pancino}, {Pauwels}, {Recio-Blanco}, {Reyl{\'e}}, {Riello}, {Rimoldini}, {Roegiers}, {Rybizki}, {Sarro}, {Siopis}, {Smith}, {Sozzetti}, {Utrilla}, {van Leeuwen}, {Abbas}, {{\'A}brah{\'a}m}, {Abreu Aramburu}, {Aerts}, {Aguado}, {ajaj}, {Aldea-Montero}, {Altavilla}, {{\'A}lvarez}, {Alves}, {Anders}, {Anderson}, {Anglada Varela}, {Antoja}, {Baines}, {Baker}, {Balaguer-N{\'u}{\~n}ez}, {Balbinot}, {Balog}, {Barache}, {Barbato}, {Barros}, {Barstow}, {Bartolom{\'e}}, {Bassilana}, {Bauchet}, {Becciani}, {Bellazzini}, {Berihuete}, {Bernet}, {Bertone}, {Bianchi}, {Binnenfeld}, {Blanco-Cuaresma}, {Blazere}, {Boch}, {Bombrun}, {Bossini}, {Bouquillon}, {Bragaglia}, {Bramante}, {Breedt},
  {Bressan}, {Brouillet}, {Brugaletta}, {Bucciarelli}, {Burlacu}, {Butkevich}, {Buzzi}, {Caffau}, {Cancelliere}, {Cantat-Gaudin}, {Carballo}, {Carlucci}, {Carnerero}, {Carrasco}, {Casamiquela}, {Castellani}, {Castro-Ginard}, {Chaoul}, {Charlot}, {Chemin}, {Chiaramida}, {Chiavassa}, {Chornay}, {Comoretto}, {Contursi}, {Cooper}, {Cornez}, {Cowell}, {Crifo}, {Cropper}, {Crosta}, {Crowley}, {Dafonte}, {Dapergolas}, {David}, {David}, {de Laverny}, {De Luise}, {De March}, {De Ridder}, {de Souza}, {de Torres}, {del Peloso}, {del Pozo}, {Delbo}, {Delgado}, {Delisle}, {Demouchy}, {Dharmawardena}, {Di Matteo}, {Diakite}, {Diener}, {Distefano}, {Dolding}, {Edvardsson}, {Enke}, {Fabre}, {Fabrizio}, {Faigler}, {Fedorets}, {Fernique}, {Fienga}, {Figueras}, {Fournier}, {Fouron}, {Fragkoudi}, {Gai}, {Garcia-Gutierrez}, {Garcia-Reinaldos}, {Garc{\'\i}a-Torres}, {Garofalo}, {Gavel}, {Gavras}, {Gerlach}, {Geyer}, {Giacobbe}, {Gilmore}, {Girona}, {Giuffrida}, {Gomel}, {Gomez}, {Gonz{\'a}lez-N{\'u}{\~n}ez},
  {Gonz{\'a}lez-Santamar{\'\i}a}, {Gonz{\'a}lez-Vidal}, {Granvik}, {Guillout}, {Guiraud}, {Guti{\'e}rrez-S{\'a}nchez}, {Guy}, {Hatzidimitriou}, {Hauser}, {Haywood}, {Helmer}, {Helmi}, {Sarmiento}, {Hidalgo}, {Hilger}, {H{\l}adczuk}, {Hobbs}, {Holland}, {Huckle}, {Jardine}, {Jasniewicz}, {Jean-Antoine Piccolo}, {Jim{\'e}nez-Arranz}, {Jorissen}, {Juaristi Campillo}, {Julbe}, {Karbevska}, {Kervella}, {Khanna}, {Kontizas}, {Kordopatis}, {Korn}, {K{\'o}sp{\'a}l}, {Kostrzewa-Rutkowska}, {Kruszy{\'n}ska}, {Kun}, {Laizeau}, {Lambert}, {Lanza}, {Lasne}, {Le Campion}, {Lebreton}, {Lebzelter}, {Leccia}, {Leclerc}, {Lecoeur-Taibi}, {Liao}, {Licata}, {Lindstr{\o}m}, {Lister}, {Livanou}, {Lobel}, {Lorca}, {Loup}, {Madrero Pardo}, {Magdaleno Romeo}, {Managau}, {Mann}, {Manteiga}, {Marchant}, {Marconi}, {Marcos}, {Marcos Santos}, {Mar{\'\i}n Pina}, {Marinoni}, {Marocco}, {Marshall}, {Martin Polo}, {Mart{\'\i}n-Fleitas}, {Marton}, {Mary}, {Masip}, {Massari}, {Mastrobuono-Battisti}, {Mazeh}, {McMillan}, {Messina}, {Michalik},
  {Millar}, {Mints}, {Molina}, {Molinaro}, {Moln{\'a}r}, {Monari}, {Mongui{\'o}}, {Montegriffo}, {Montero}, {Mor}, {Mora}, {Morbidelli}, {Morel}, {Morris}, {Muraveva}, {Murphy}, {Musella}, {Nagy}, {Noval}, {Oca{\~n}a}, {Ogden}, {Ordenovic}, {Osinde}, {Pagani}, {Pagano}, {Palaversa}, {Palicio}, {Pallas-Quintela}, {Panahi}, {Payne-Wardenaar}, {Pe{\~n}alosa Esteller}, {Penttil{\"a}}, {Pichon}, {Piersimoni}, {Pineau}, {Plachy}, {Plum}, {Poggio}, {Pr{\v{s}}a}, {Pulone}, {Racero}, {Ragaini}, {Rainer}, {Raiteri}, {Rambaux}, {Ramos}, {Ramos-Lerate}, {Re Fiorentin}, {Regibo}, {Richards}, {Rios Diaz}, {Ripepi}, {Riva}, {Rix}, {Rixon}, {Robichon}, {Robin}, {Robin}, {Roelens}, {Rogues}, {Rohrbasser}, {Romero-G{\'o}mez}, {Rowell}, {Royer}, {Ruz Mieres}, {Rybicki}, {Sadowski}, {S{\'a}ez N{\'u}{\~n}ez}, {Sagrist{\`a} Sell{\'e}s}, {Sahlmann}, {Salguero}, {Samaras}, {Sanchez Gimenez}, {Sanna}, {Santove{\~n}a}, {Sarasso}, {Schultheis}, {Sciacca}, {Segol}, {Segovia}, {S{\'e}gransan}, {Semeux}, {Shahaf}, {Siddiqui}, {Siebert},
  {Siltala}, {Silvelo}, {Slezak}, {Slezak}, {Smart}, {Snaith}, {Solano}, {Solitro}, {Souami}, {Souchay}, {Spagna}, {Spina}, {Spoto}, {Steele}, {Steidelm{\"u}ller}, {Stephenson}, {S{\"u}veges}, {Surdej}, {Szabados}, {Szegedi-Elek}, {Taris}, {Taylor}, {Teixeira}, {Tolomei}, {Tonello}, {Torra}, {Torra}, {Torralba Elipe}, {Trabucchi}, {Tsounis}, {Turon}, {Ulla}, {Unger}, {Vaillant}, {van Dillen}, {van Reeven}, {Vanel}, {Vecchiato}, {Viala}, {Vicente}, {Voutsinas}, {Weiler}, {Wevers}, {Wyrzykowski}, {Yoldas}, {Yvard}, {Zhao}, {Zorec}, {Zucker}, \& {Zwitter}}]{gaiaDR3_2023}
{Gaia Collaboration}, {Vallenari}, A., {Brown}, A.~G.~A., {et~al.} 2023, \aap, 674, A1

\bibitem[{{Gandhi} {et~al.}(2022){Gandhi}, {Kesseli}, {Snellen}, {Brogi}, {Wardenier}, {Parmentier}, {Welbanks}, \& {Savel}}]{gandhi2022}
{Gandhi}, S., {Kesseli}, A., {Snellen}, I., {et~al.} 2022, \mnras, 515, 749

\bibitem[{{Gandhi} {et~al.}(2023){Gandhi}, {Kesseli}, {Zhang}, {Louca}, {Snellen}, {Brogi}, {Miguel}, {Casasayas-Barris}, {Pelletier}, {Landman}, {Maguire}, \& {Gibson}}]{gandhi2023}
{Gandhi}, S., {Kesseli}, A., {Zhang}, Y., {et~al.} 2023, \aj, 165, 242

\bibitem[{{Gelman} \& {Rubin}(1992)}]{gelman1992}
{Gelman}, A. \& {Rubin}, D.~B. 1992, StaSc, 7, 457

\bibitem[{{Giacobbe} {et~al.}(2021){Giacobbe}, {Brogi}, {Gandhi}, {Cubillos}, {Bonomo}, {Sozzetti}, {Fossati}, {Guilluy}, {Carleo}, {Rainer}, {Harutyunyan}, {Borsa}, {Pino}, {Nascimbeni}, {Benatti}, {Biazzo}, {Bignamini}, {Chubb}, {Claudi}, {Cosentino}, {Covino}, {Damasso}, {Desidera}, {Fiorenzano}, {Ghedina}, {Lanza}, {Leto}, {Maggio}, {Malavolta}, {Maldonado}, {Micela}, {Molinari}, {Pagano}, {Pedani}, {Piotto}, {Poretti}, {Scandariato}, {Yurchenko}, {Fantinel}, {Galli}, {Lodi}, {Sanna}, \& {Tozzi}}]{giacobbe2021}
{Giacobbe}, P., {Brogi}, M., {Gandhi}, S., {et~al.} 2021, Nat, 592, 205

\bibitem[{{Gibson} {et~al.}(2020){Gibson}, {Merritt}, {Nugroho}, {Cubillos}, {de Mooij}, {Mikal-Evans}, {Fossati}, {Lothringer}, {Nikolov}, {Sing}, {Spake}, {Watson}, \& {Wilson}}]{gibson2020}
{Gibson}, N.~P., {Merritt}, S., {Nugroho}, S.~K., {et~al.} 2020, \mnras, 493, 2215

\bibitem[{{Hargreaves} {et~al.}(2020){Hargreaves}, {Gordon}, {Rey}, {Nikitin}, {Tyuterev}, {Kochanov}, \& {Rothman}}]{ch4_hitemp}
{Hargreaves}, R.~J., {Gordon}, I.~E., {Rey}, M., {et~al.} 2020, \apjs, 247, 55

\bibitem[{{Heng} \& {Showman}(2015)}]{heng2015}
{Heng}, K. \& {Showman}, A.~P. 2015, AREPS, 43, 509

\bibitem[{{Hinkel} {et~al.}(2014){Hinkel}, {Timmes}, {Young}, {Pagano}, \& {Turnbull}}]{Hinkel2014}
{Hinkel}, N.~R., {Timmes}, F.~X., {Young}, P.~A., {Pagano}, M.~D., \& {Turnbull}, M.~C. 2014, \aj, 148, 54

\bibitem[{{H{\o}g} {et~al.}(2000){H{\o}g}, {Fabricius}, {Makarov}, {Urban}, {Corbin}, {Wycoff}, {Bastian}, {Schwekendiek}, \& {Wicenec}}]{hog2000}
{H{\o}g}, E., {Fabricius}, C., {Makarov}, V.~V., {et~al.} 2000, \aap, 355, L27

\bibitem[{{Huang} {et~al.}(2017){Huang}, Schwenke, Freedman, \& Lee}]{co2_ames}
{Huang}, X., Schwenke, D.~W., Freedman, R.~S., \& Lee, T.~J. 2017, \jqsrt, 203, 224, hITRAN2016 Special Issue

\bibitem[{{Johns} {et~al.}(2019){Johns}, {Reed}, {Rodriguez}, {Pepper}, {Stassun}, {Penev}, {Gaudi}, {Labadie-Bartz}, {Fulton}, {Quinn}, {Eastman}, {Ciardi}, {Hirsch}, {Stevens}, {Stevens}, {Oberst}, {Cohen}, {Jensen}, {Benni}, {Villanueva}, {Murawski}, {Bieryla}, {Latham}, {Vanaverbeke}, {Dubois}, {Rau}, {Logie}, {Rauenzahn}, {Wittenmyer}, {Zambelli}, {Bayliss}, {Beatty}, {Collins}, {Col{\'o}n}, {Curtis}, {Evans}, {Gregorio}, {James}, {Depoy}, {Johnson}, {Joner}, {Kasper}, {Khakpash}, {Kielkopf}, {Kuhn}, {Lund}, {Manner}, {Marshall}, {McLeod}, {Penny}, {Relles}, {Siverd}, {Stephens}, {Stockdale}, {Tan}, {Trueblood}, {Trueblood}, \& {Yao}}]{johns2019}
{Johns}, D., {Reed}, P.~A., {Rodriguez}, J.~E., {et~al.} 2019, \aj, 158, 78

\bibitem[{Kasper {et~al.}(2023)Kasper, Bean, Line, Seifahrt, Brady, Lothringer, Pino, Fu, Pelletier, Stürmer, Benneke, Brogi, \& Désert}]{kasper2023}
Kasper, D., Bean, J.~L., Line, M.~R., {et~al.} 2023, \aj, 165, 7

\bibitem[{Kasper {et~al.}(2021)Kasper, Bean, Line, Seifahrt, Stürmer, Pino, Désert, \& Brogi}]{kasper2021}
Kasper, D., Bean, J.~L., Line, M.~R., {et~al.} 2021, ApJL, 921, L18

\bibitem[{{Kokori} {et~al.}(2023){Kokori}, {Tsiaras}, {Edwards}, {Jones}, {Pantelidou}, {Tinetti}, {Bewersdorff}, {Iliadou}, {Jongen}, {Lekkas}, {Nastasi}, {Poultourtzidis}, {Sidiropoulos}, {Walter}, {W{\"u}nsche}, {Abraham}, {Agnihotri}, {Albanesi}, {Arce-Mansego}, {Arnot}, {Audejean}, {Aumasson}, {Bachschmidt}, {Baj}, {Barroy}, {Belinski}, {Bennett}, {Benni}, {Bernacki}, {Betti}, {Biagini}, {Bosch}, {Brandebourg}, {Br{\'a}t}, {Bretton}, {Brincat}, {Brouillard}, {Bruzas}, {Bruzzone}, {Buckland}, {Cal{\'o}}, {Campos}, {Carre{\~n}o}, {Carrion Rodrigo}, {Casali}, {Casalnuovo}, {Cataneo}, {Chang}, {Changeat}, {Chowdhury}, {Ciantini}, {Cilluffo}, {Coliac}, {Conzo}, {Correa}, {Coulon}, {Crouzet}, {Crow}, {Curtis}, {Daniel}, {Dauchet}, {Dawes}, {Deldem}, {Deligeorgopoulos}, {Dransfield}, {Dymock}, {Eenm{\"a}e}, {Esseiva}, {Evans}, {Falco}, {Farf{\'a}n}, {Fern{\'a}ndez-Laj{\'u}s}, {Ferratfiat}, {Ferreira}, {Ferretti}, {Fio{\l}ka}, {Fowler}, {Futcher}, {Gabellini}, {Gainey}, {Gaitan}, {Gajdo{\v{s}}},
  {Garc{\'\i}a-S{\'a}nchez}, {Garlitz}, {Gillier}, {Gison}, {Gonzales}, {Gorshanov}, {Grau Horta}, {Grivas}, {Guerra}, {Guillot}, {Haswell}, {Haymes}, {Hentunen}, {Hills}, {Hose}, {Humbert}, {Hurter}, {Hynek}, {Irzyk}, {Jacobsen}, {Jannetta}, {Johnson}, {J{\'o}{\'z}wik-Wabik}, {Kaeouach}, {Kang}, {Kiiskinen}, {Kim}, {Kivila}, {Koch}, {Kolb}, {Ku{\v{c}}{\'a}kov{\'a}}, {Lai}, {Laloum}, {Lasota}, {Lewis}, {Liakos}, {Libotte}, {Lomoz}, {Lopresti}, {Majewski}, {Malcher}, {Mallonn}, {Mannucci}, {Marchini}, {Mari}, {Marino}, {Marino}, {Mario}, {Marquette}, {Mart{\'\i}nez-Bravo}, {Ma{\v{s}}ek}, {Matassa}, {Michel}, {Michelet}, {Miller}, {Miny}, {Molina}, {Mollier}, {Monteleone}, {Montigiani}, {Morales-Aimar}, {Mortari}, {Morvan}, {Mugnai}, {Murawski}, {Naponiello}, {Naudin}, {Naves}, {N{\'e}el}, {Neito}, {Neveu}, {Noschese}, {{\"O}{\u{g}}men}, {Ohshima}, {Orbanic}, {Pace}, {Pantacchini}, {Paschalis}, {Pereira}, {Peretto}, {Perroud}, {Phillips}, {Pintr}, {Pioppa}, {Plazas}, {Poelarends}, {Popowicz}, {Purcell},
  {Quinn}, {Raetz}, {Rees}, {Regembal}, {Rocchetto}, {Rocci}, {Rockenbauer}, {Roth}, {Rousselot}, {Rubia}, {Ruocco}, {Russo}, {Salisbury}, {Salvaggio}, {Santos}, {Savage}, {Scaggiante}, {Sedita}, {Shadick}, {Silva}, {Sioulas}, {{\v{S}}koln{\'\i}k}, {Smith}, {Smolka}, {Solmaz}, {Stanbury}, {Stouraitis}, {Tan}, {Theusner}, {Thurston}, {Tifner}, {Tomacelli}, {Tomatis}, {Trnka}, {Tyl{\v{s}}ar}, {Valeau}, {Vignes}, {Villa}, {Vives Sureda}, {Vora}, {Vra{\v{s}}t'{\'a}k}, {Walliang}, {Wenzel}, {Wright}, {Zambelli}, {Zhang}, \& {Z{\'\i}bar}}]{kokori2023}
{Kokori}, A., {Tsiaras}, A., {Edwards}, B., {et~al.} 2023, \apjs, 265, 4

\bibitem[{{Kozai}(1962)}]{kozai1962}
{Kozai}, Y. 1962, \aj, 67, 591

\bibitem[{{Lecavelier Des Etangs} {et~al.}(2008){Lecavelier Des Etangs}, {Pont}, {Vidal-Madjar}, \& {Sing}}]{lecavelier2008}
{Lecavelier Des Etangs}, A., {Pont}, F., {Vidal-Madjar}, A., \& {Sing}, D. 2008, \aap, 481, L83

\bibitem[{{Line} {et~al.}(2021){Line}, {Brogi}, {Bean}, {Gandhi}, {Zalesky}, {Parmentier}, {Smith}, {Mace}, {Mansfield}, {Kempton}, {Fortney}, {Shkolnik}, {Patience}, {Rauscher}, {D{\'e}sert}, \& {Wardenier}}]{line2021}
{Line}, M.~R., {Brogi}, M., {Bean}, J.~L., {et~al.} 2021, Nat, 598, 580

\bibitem[{{Lodders}(2010)}]{lodders2010}
{Lodders}, K. 2010, ApSSP, 16, 379

\bibitem[{MacDonald {et~al.}(2020)MacDonald, Goyal, \& Lewis}]{macdonald2020}
MacDonald, R.~J., Goyal, J.~M., \& Lewis, N.~K. 2020, ApJL, 893, L43

\bibitem[{{Madhusudhan}(2012)}]{madhusudhan2012}
{Madhusudhan}, N. 2012, \apj, 758, 36

\bibitem[{{Madhusudhan}(2019)}]{madhusudhan2019}
{Madhusudhan}, N. 2019, \araa, 57, 617

\bibitem[{{Madhusudhan} {et~al.}(2014){Madhusudhan}, {Knutson}, {Fortney}, \& {Barman}}]{madhusudhan2014}
{Madhusudhan}, N., {Knutson}, H., {Fortney}, J.~J., \& {Barman}, T. 2014, in Protostars and Planets VI, ed. H.~{Beuther}, R.~S. {Klessen}, C.~P. {Dullemond}, \& T.~{Henning}, 739--762

\bibitem[{Madhusudhan {et~al.}(2020)Madhusudhan, Nixon, Welbanks, Piette, \& Booth}]{madhusudhan2020}
Madhusudhan, N., Nixon, M.~C., Welbanks, L., Piette, A. A.~A., \& Booth, R.~A. 2020, ApJL, 891, L7

\bibitem[{Mansfield {et~al.}(2024)Mansfield, Line, Wardenier, Brogi, Bean, Beltz, Smith, Zalesky, Batalha, Kempton, Montet, Owen, Plavchan, \& Rauscher}]{mansfield2024}
Mansfield, M.~W., Line, M.~R., Wardenier, J.~P., {et~al.} 2024, \aj, 168, 14

\bibitem[{{Molli{\`e}re} {et~al.}(2019){Molli{\`e}re}, {Wardenier}, {van Boekel}, {Henning}, {Molaverdikhani}, \& {Snellen}}]{molliere2019}
{Molli{\`e}re}, P., {Wardenier}, J.~P., {van Boekel}, R., {et~al.} 2019, \aap, 627, A67

\bibitem[{Mordasini {et~al.}(2016)Mordasini, van Boekel, Mollière, Henning, \& Benneke}]{mordasini2016}
Mordasini, C., van Boekel, R., Mollière, P., Henning, T., \& Benneke, B. 2016, \apj, 832, 41

\bibitem[{{Moses}(2014)}]{moses2014}
{Moses}, J.~I. 2014, RSPTA, 372, 20130073

\bibitem[{{Noll} {et~al.}(2012){Noll}, {Kausch}, {Barden}, {Jones}, {Szyszka}, {Kimeswenger}, \& {Vinther}}]{noll2012}
{Noll}, S., {Kausch}, W., {Barden}, M., {et~al.} 2012, \aap, 543, A92

\bibitem[{{Nortmann} {et~al.}(2025){Nortmann}, {Lesjak}, {Yan}, {Cont}, {Czesla}, {Lavail}, {Rains}, {Nagel}, {Boldt-Christmas}, {Hatzes}, {Reiners}, {Piskunov}, {Kochukhov}, {Heiter}, {Shulyak}, {Rengel}, \& {Seemann}}]{nortmann2025}
{Nortmann}, L., {Lesjak}, F., {Yan}, F., {et~al.} 2025, \aap, 693, A213

\bibitem[{{{\"O}berg} {et~al.}(2011){{\"O}berg}, {Murray-Clay}, \& {Bergin}}]{oberg2011}
{{\"O}berg}, K.~I., {Murray-Clay}, R., \& {Bergin}, E.~A. 2011, \apjl, 743, L16

\bibitem[{Pacetti {et~al.}(2022)Pacetti, Turrini, Schisano, Molinari, Fonte, Politi, Hennebelle, Klessen, Testi, \& Lebreuilly}]{pacetti2022}
Pacetti, E., Turrini, D., Schisano, E., {et~al.} 2022, \apj, 937, 36

\bibitem[{{Paxton} {et~al.}(2015){Paxton}, {Marchant}, {Schwab}, {Bauer}, {Bildsten}, {Cantiello}, {Dessart}, {Farmer}, {Hu}, {Langer}, {Townsend}, {Townsley}, \& {Timmes}}]{2015ApJS..220...15P}
{Paxton}, B., {Marchant}, P., {Schwab}, J., {et~al.} 2015, \apjs, 220, 15

\bibitem[{{Pepe} {et~al.}(2002){Pepe}, {Mayor}, {Galland}, {Naef}, {Queloz}, {Santos}, {Udry}, \& {Burnet}}]{pepe2002}
{Pepe}, F., {Mayor}, M., {Galland}, F., {et~al.} 2002, \aap, 388, 632

\bibitem[{Pepper {et~al.}(2007)Pepper, Pogge, DePoy, Marshall, Stanek, Stutz, Poindexter, Siverd, O’Brien, Trueblood, \& Trueblood}]{pepper2007}
Pepper, J., Pogge, R.~W., DePoy, D.~L., {et~al.} 2007, \pasp, 119, 923

\bibitem[{{Pino} {et~al.}(2022){Pino}, {Brogi}, {D{\'e}sert}, {Nascimbeni}, {Bonomo}, {Rauscher}, {Basilicata}, {Biazzo}, {Bignamini}, {Borsa}, {Claudi}, {Covino}, {Di Mauro}, {Guilluy}, {Maggio}, {Malavolta}, {Micela}, {Molinari}, {Molinaro}, {Montalto}, {Nardiello}, {Pedani}, {Piotto}, {Poretti}, {Rainer}, {Scandariato}, {Sicilia}, \& {Sozzetti}}]{pino2022}
{Pino}, L., {Brogi}, M., {D{\'e}sert}, J.~M., {et~al.} 2022, \aap, 668, A176

\bibitem[{Polyansky {et~al.}(2018)Polyansky, Kyuberis, Zobov, Tennyson, Yurchenko, \& Lodi}]{h2o_pokazatel}
Polyansky, O.~L., Kyuberis, A.~A., Zobov, N.~F., {et~al.} 2018, \mnras, 480, 2597

\bibitem[{Rainer {et~al.}(2018)Rainer, Harutyunyan, Carleo, Oliva, Benatti, Bignamini, Claudi, Gonzalez-Alvarez, Sanna, Ghedina, Micela, Molinari, Tozzi, Baffa, Baruffolo, Buchschacher, Cecconi, Cosentino, Falcini, Fantinel, Fini, Galli, Ghinassi, Giani, Gonzalez, Gonzalez, Gratton, Guerra, Diaz, Hernandez, Iuzzolino, Lodi, Malavolta, Maldonado, Origlia, Ventura, Puglisi, Riverol, Riverol, Juan, Scuderi, Seeman, Sozzetti, \& Sozzi}]{rainer2018}
Rainer, M., Harutyunyan, A., Carleo, I., {et~al.} 2018, in Ground-based and Airborne Instrumentation for Astronomy VII, ed. C.~J. Evans, L.~Simard, \& H.~Takami, Vol. 10702, International Society for Optics and Photonics (SPIE), 1070266

\bibitem[{{Ram{\'\i}rez} {et~al.}(2014){Ram{\'\i}rez}, {Mel{\'e}ndez}, {Bean}, {Asplund}, {Bedell}, {Monroe}, {Casagrande}, {Schirbel}, {Dreizler}, {Teske}, {Tucci Maia}, {Alves-Brito}, \& {Baumann}}]{Ramirez2014}
{Ram{\'\i}rez}, I., {Mel{\'e}ndez}, J., {Bean}, J., {et~al.} 2014, \aap, 572, A48

\bibitem[{Rothman {et~al.}(2010)Rothman, Gordon, Barber, Dothe, Gamache, Goldman, Perevalov, Tashkun, \& Tennyson}]{co_hitemp}
Rothman, L., Gordon, I., Barber, R., {et~al.} 2010, \jqsrt, 111, 2139, xVIth Symposium on High Resolution Molecular Spectroscopy (HighRus-2009)

\bibitem[{{Schneider} \& {Bitsch}(2021)}]{schneider2021}
{Schneider}, A.~D. \& {Bitsch}, B. 2021, \aap, 654, A71

\bibitem[{{Showman} \& {Polvani}(2011)}]{showman2011}
{Showman}, A.~P. \& {Polvani}, L.~M. 2011, \apj, 738, 71

\bibitem[{Sing(2018)}]{sing2018}
Sing, D.~K. 2018, Observational Techniques with Transiting Exoplanetary Atmospheres, ed. V.~Bozza, L.~Mancini, \& A.~Sozzetti (Cham: Springer International Publishing), 3--48

\bibitem[{{Sneden} {et~al.}(2012){Sneden}, {Bean}, {Ivans}, {Lucatello}, \& {Sobeck}}]{moog}
{Sneden}, C., {Bean}, J., {Ivans}, I., {Lucatello}, S., \& {Sobeck}, J. 2012, {MOOG: LTE line analysis and spectrum synthesis}, Astrophysics Source Code Library, record ascl:1202.009

\bibitem[{{Snellen} {et~al.}(2010){Snellen}, {de Kok}, {de Mooij}, \& {Albrecht}}]{snellen2010}
{Snellen}, I. A.~G., {de Kok}, R.~J., {de Mooij}, E. J.~W., \& {Albrecht}, S. 2010, Nat, 465, 1049

\bibitem[{{Sousa} {et~al.}(2015){Sousa}, {Santos}, {Adibekyan}, {Delgado-Mena}, \& {Israelian}}]{sousa2015}
{Sousa}, S.~G., {Santos}, N.~C., {Adibekyan}, V., {Delgado-Mena}, E., \& {Israelian}, G. 2015, \aap, 577, A67

\bibitem[{{Ter Braak}(2006)}]{terbraak2006}
{Ter Braak}, C. J.~F. 2006, Statistics and Computing, 16, 239

\bibitem[{{Turrini} {et~al.}(2021){Turrini}, {Schisano}, {Fonte}, {Molinari}, {Politi}, {Fedele}, {Pani{\'c}}, {Kama}, {Changeat}, \& {Tinetti}}]{turrini2021}
{Turrini}, D., {Schisano}, E., {Fonte}, S., {et~al.} 2021, \apj, 909, 40

\bibitem[{{Wardenier} {et~al.}(2023){Wardenier}, {Parmentier}, {Line}, \& {Lee}}]{wardenier2023}
{Wardenier}, J.~P., {Parmentier}, V., {Line}, M.~R., \& {Lee}, E. K.~H. 2023, \mnras, 525, 4942

\bibitem[{{Wardenier} {et~al.}(2024){Wardenier}, {Parmentier}, {Line}, {Weiner Mansfield}, {Tan}, {Tsai}, {Bean}, {Birkby}, {Brogi}, {D{\'e}sert}, {Gandhi}, {Lee}, {Levens}, {Pino}, \& {Smith}}]{wardenier2024}
{Wardenier}, J.~P., {Parmentier}, V., {Line}, M.~R., {et~al.} 2024, \pasp, 136, 084403

\bibitem[{{Xue} {et~al.}(2024){Xue}, {Bean}, {Zhang}, {Welbanks}, {Lunine}, \& {August}}]{xue2024}
{Xue}, Q., {Bean}, J.~L., {Zhang}, M., {et~al.} 2024, \apjl, 963, L5

\end{thebibliography}
\end{document}